\documentclass[12pt]{iopart}

\usepackage{cite}
\usepackage{graphicx,color}
\usepackage{amssymb}
\usepackage{amsthm}
\newcommand{\ket}[1]{|{#1}\rangle}
\newcommand{\bra}[1]{\langle{#1}|}
\newcommand{\Det}{\mathop{\mathrm{Det}}}
\newcommand{\mod}{\mathrm{mod}}
\newcommand{\Poly}{\mathop{\mathrm{Poly}}\nolimits}
\newcommand{\Supp}{\mathop{\mathrm{Supp}}}
\newcommand{\Ker}{\mathop{\mathrm{Ker}}}
\newcommand{\Span}{\mathop{\mathrm{Span}}}
\newtheorem{lemma}{Lemma}
\newtheorem{theo}{Theorem}
\newtheorem{cor}{Corollary}
\newtheorem{defi}{Definition}
\newtheorem{prop}{Proposition}
\newtheorem{remark}{Remark}
\newcommand{\Proof}{\noindent\textbf{Proof.} } 
\newcommand{\Proofa}[1]{\noindent\textbf{#1.} } 

\newcommand{\red}{\color{black}}
\definecolor{dgreen}{rgb}{0,0,0}

\newcommand{\blue}{\color{black}}

\begin{document}
\title{Ergodic and Mixing Quantum Channels in Finite Dimensions}
\author{D. Burgarth$^1$, G. Chiribella$^{2}$, V. Giovannetti$^{3}$, P. Perinotti$^{4}$, and  K. Yuasa$^{5}$}
\address{$^1$Institute of Mathematics and Physics, Aberystwyth University, Physical Sciences Building, Penglais Campus, SY23 3BZ Aberystwyth, United Kingdom}
\address{$^2$Center for Quantum Information, Institute for Interdisciplinary Information Sciences, Tsinghua University, Beijing 100084, China}
\address{$^3$NEST, Scuola Normale Superiore and Istituto Nanoscienze-CNR,  piazza dei Cavalieri 7, I-56126 Pisa, Italy}     
\address{$^4$Dipartimento di Fisica, Universit\`a degli Studi di Pavia and INFN sezione di Pavia, via Bassi 6, I-27100 Pavia, Italy}
\address{$^5$Department of Physics, Waseda University, Tokyo 169-8555, Japan}

\date \today

\begin{abstract}
The paper provides a systematic characterization of quantum ergodic and mixing channels in finite dimensions and a discussion of their structural properties. In particular, we {\blue discuss} ergodicity in the general case where the fixed point of the channel is not a full-rank (faithful) density matrix.  Notably, we show that ergodicity is stable under randomizations, namely that every random mixture of an ergodic channel with a generic channel is still ergodic. In addition, we prove several conditions under which ergodicity can be promoted to the stronger property of mixing. Finally, exploiting a suitable correspondence between quantum channels and generators of quantum dynamical semigroups, we extend our results to the realm of continuous-time quantum evolutions, providing a characterization of ergodic Lindblad generators and showing that they are dense in the set of all possible generators. 
\end{abstract}
\pacs{}
\maketitle

\section{Introduction}
In the study of signal processing, a stochastic process is said to be
ergodic if its statistical properties can be deduced from a single,
sufficiently long realization of the process.  Ergodicity
plays a fundamental role in the study of the irreversible dynamics
associated with the relaxation to the thermal equilibrium
{\blue \cite{STREATER,BRPE,BEFLO,ALICKI,oseledets,sarymsakov,HOLEVObook,wielandt}}.  In quantum mechanics the evolution
of an open system, interacting with an external (initially
uncorrelated) environment, is fully characterized in terms of special
linear maps ${\cal M}$ (known as \emph{quantum channels}) operating on the space of density matrices $\rho$
of the system of interest, under certain structural constraints
(quantum channels and all their extensions should preserve
the positivity and the trace of the operators on which they act upon,
see e.g.\ \cite{REV,KEYL,BENSHOR,WOLF}).  
{\blue  In this framework the rigorous definition of quantum ergodic channels 
can be traced back to a series  of works that appeared in the late 70's, which set the proper mathematical background and
clarified the main aspects of the problem.
For a review on the subject  see e.g.\ Refs.~\cite{oseledets,HOLEVObook,sarymsakov,SPOHN}. 
 A convenient  definition of ergodic quantum channels  can be given as follows: 
  a quantum  channel $\mathcal M$}  is ergodic if and only if it  admits a unique fixed point in the space of density matrices, that is, if there is only one density matrix $\rho_*$  that  is unaltered by the
action of ${\cal M}$ {\blue \cite{NJP,RAGI,RAGI2,STREATER}}. 
 The rationale behind
such formulation   is clear when we consider the discrete trajectories
associated with the evolution of a generic input state $\rho$,
evolving under iterated applications of the transformation ${\cal
  M}$: in this case, the mean value of a generic observable  $A$, averaged over the trajectories, converges
asymptotically to the  expectation value $\Tr[A \rho_*]$ of the observable on
the fixed point {\blue\cite{oseledets,EVANS,MOROZOVA}.\footnote{{\blue It is worth observing that elsewhere (e.g.\ Refs.~\cite{HOLEVObook,oseledets,EVANS,MOROZOVA,DAVIES})
 the convergency  of the mean values averaged over the system trajectory   
 is adopted as the formal way for defining ergodicity, while  the existence of a
unique fixed point is associated with the notion of {\em irreducibility} of the map (notice however that some Authors use this term only to indicate ergodic channels with faithful fixed points, see e.g.\ \cite{WOLF}, while other use the name {\em strong irreducibility} to identify  channels which are mixing and have faithful fixed points, see e.g.\ \cite{wielandt}). Since for quantum channels the two properties  are equivalent our choice of not distinguishing them is not misleading.}}  
A related, but stronger property of an ergodic quantum channel, is the ability of transforming a generic input state into the fixed point $\rho_*$  after a sufficiently large number of repeated applications.
In the context of dynamical semigroups~\cite{SPOHN}
 this property is often called {\it relaxing}, here instead we  follow the notation of~\cite{NJP} and dubbed  it {\it mixing}.}

Beyond the study of relaxation processes, ergodicity  and mixing have found important applications in several fields
of quantum information theory. Most notably in quantum control \cite{BURK,MEDIATED,MEDIATED2,CONT,CONT2,CONT3,CONT4},  quantum estimation \cite{GUTA}, quantum communication \cite{COMM,COMM2}, and
in the study of efficient
 tensorial representation of critical many-body quantum systems \cite{matrixprod,MERA,MERA2,MERA3,MERA4,MERA5}.
 {\blue A detailed analysis of ergodicity in the quantum domain appears hence to be mandatory.  
   Here we contribute to this goal by providing a systematic characterization of the structural properties of ergodic and mixing channels on finite-dimensional quantum systems. 
   The results are presented in a systematic way, starting from the case of general channels and then specializing to particular classes of channels (such as random-unitary channels). For completeness we also include  some alternative proofs of existing results, which come out quite naturally in our approach and which, unlike most results in the previous literature,
   are  derived without making explicit use of the faithfulness of the fixed point unless such assumption is strictly necessary. 
This allows us to guide the reader through some key aspects of the structure of ergodic and mixing
 channels, presenting both new and old results in a compact, self-contained 
form (still providing, whenever possible,  proper references to the original works).}  {\red A further advantage of our presentation with respect to the original literature of the 70's  is that here all proofs are elementary, due to our focus on \emph{finite dimensions}, and often what is presented here as a single theorem (with a rather straightforward proof) was the subject of a full paper in the original von Neumann algebraic setting, thus preventing a comprehensive overlook on the subject.

{\em Summary of the main results.}   The paper starts with a characterization of ergodicity for channels without a faithful fixed point, provided in Theorem \ref{theo:ergogen}: a quantum channel is ergodic if and only if it possesses a minimal invariant subspace (a subspace of the Hilbert space that is left invariant by the action of the channel and does not contain any smaller, non-trivial subspace with the same property).     
 Later, this characterization is used to prove a fundamental property of ergodic channels,
namely the fact that \emph{ergodicity is stable under randomizations} (Theorem \ref{theo:ergoconv}).  Precisely, we show that  a convex combination of an ergodic channel with a generic (not necessarily ergodic) channel
yields a new transformation which is always ergodic.   In addition, we show that the fixed point of the new ergodic map is related
with the fixed point of the original ergodic channel via a convex combination.  These  results 
 extend to the case of ergodic channels a property that was previously known to hold for the restricted subset of mixing channels \cite{MEDIATED,MEDIATED2,MEDI1}.  
 In addition, we provide a series of conditions under which ergodicity can be upgraded to the stronger property of mixing: for example, we show that, rather counterintuitively, convex combinations of ergodic channels with the identity channel are always mixing.   Another remarkable property is that a random-unitary channel  $\mathcal M$ (a channel that is a convex combination of unitaries) in dimension $d$ is mixing if and only if the channel $\mathcal M^d$ obtained by applying $\mathcal M$ on the system $d$ times is ergodic.  
 Finally, we discuss the case of quantum dynamical semigroups~\cite{SPOHN}.
In this context we show that the ergodicity of a quantum dynamical semigroup is equivalent to the ergodicity of a suitable quantum channel (Theorem \ref{theo:ergogroup}) and, exploiting the properties of the latter, 
we prove a stability property under randomization of the
 generators of ergodic quantum semigroups  (Theorem \ref{convexitysemigroup}). }

The paper is organized as follows. 
We start in Section~\ref{sec:1} by reviewing some basic definitions and properties. {\blue In  Section~\ref{section3} 
we  provide a characterization of ergodicity in terms of the invariant subspaces of the channel and discuss how ergodicity and mixing properties of the latter are connected with analogous properties of its adjoint map. 
The convexity properties of ergodic channel are analyzed in Section~\ref{convex:SEC}.  Here we provide two alternative proofs of the stability of ergodicity under randomization,
and analyzing the relations between  ergodicity and mixing, 
 we show that a generic convex combination of the identity map with an
ergodic channel is mixing.  Section~\ref{faithfulsec} is instead specialized on the case of 
channels which admit faithful fixed points, providing a characterization of the peripheral  spectrum of the maps and introducing
a necessary and sufficient condition which ergodic maps have to fulfill in order to be mixing.}
{\red In Section~\ref{semi:SEC}  we discuss the case of continuous-time dynamical semigroups, providing a characterization of ergodicity and using it to prove the a stability property of ergodic semigroups under randomizations.}   Final remarks are  presented in Section~\ref{sec:CON}. The paper also contains an Appendix dedicated to the more technical aspects of the proofs.

\section{Definitions and Basic Properties} \label{sec:1} 
Consider a quantum system with associated Hilbert space  ${\cal H}$ of finite dimension $d< \infty$.  In what follows we will use the symbols ${\cal B}({\cal H})$ and 
$\mathfrak{S}({\cal H})\,[\subset{\cal B}({\cal H})]$ to represent the set of linear operators on ${\cal H}$  and the set of  the density matrices, respectively.  
A quantum channel operating  on the system is then defined as a linear mapping  ${\cal M}: {\cal B}({\cal H})\rightarrow {\cal B}({\cal H})$ which is completely positive and trace-preserving (CPTP in brief). 
While  referring the reader to \cite{REV,KEYL,BENSHOR,WOLF} for an exhaustive review on the subject, we find it useful    
 to recall a few  basic properties of CPTP maps, which will be exploited  in the following.

\begin{enumerate}
\item A linear map ${\cal M}: {\cal B}({\cal H})\rightarrow {\cal B}({\cal H})$ is completely positive  if and only if  it is  possible to identify a collection of Kraus operators $\{M_i\}_{i \in  \mathsf X}$ 
which, for all $A \in {\cal B}({\cal H})$, allows us to write 
\begin{equation}\label{kraus}
{\cal M}(A) = \sum_{i\in  \mathsf{X}} M_i A M_i^\dag. 
\end{equation} 
Furthermore  ${\cal M}$  is  trace-preserving if and only if the operators $M_i$ satisfy the  normalization condition 
\begin{equation}
\sum_{i\in  \mathsf X} M_i^\dag M_i = I. \label{norm}
\end{equation}

\item  The set $\mathfrak{C}({\cal H})$ formed by  the quantum channels  on the system is closed under convex combination and multiplication, i.e.\ given ${\cal M}_1, {\cal M}_2\in \mathfrak{C}({\cal H})$  and $p\in [0,1]$, 
 the transformations defined by the mappings
 \begin{eqnarray}
A\in {\cal B}({\cal H})&\quad\to\quad p {\cal M}_1(A)   + (1-p) {\cal M}_2(A), \\  A\in {\cal B}({\cal H})&\quad\to\quad({\cal M}_1 \circ {\cal M}_2)(A):= {\cal M}_1 ({\cal M}_2(A))
\end{eqnarray}
are also elements of $\mathfrak{C}({\cal H})$.

\item Any CPTP map ${\cal M}$  is nonexpansive: when $\cal M$ is applied to a couple of input states $\rho, \sigma\in \mathfrak{S}({\cal H})$, it 
produces output density matrices ${\cal M}(\rho)$, ${\cal M}(\sigma)$  whose relative distance is not greater than the original one, i.e. 
\begin{equation}
 \| {\cal M}(\rho) -{\cal M}(\sigma)\|_1
 \le \| \rho -\sigma\|_1,
 \label{ffd}
\end{equation}
where  $\| A \|_1 := \Tr[\sqrt{A^\dag A}]$ is the trace norm. 
\end{enumerate}

Since ${\cal M}$ is a linear operator defined on a linear space of dimension $d^2$, it admits up to $d^2$ distinct (complex) eigenvalues $\lambda$
 which solve the equation
 \begin{equation}
{\cal M}(A) = \lambda A ,    \label{FF}  
\end{equation} 
for some $A \in {\cal B}({\cal H})$, $A  \neq 0$. Such eigenvalues 
  can be determined as the zeros of the associated characteristic polynomial, i.e. 
\begin{equation} \label{dffd}
{\Poly}^{({\cal M})}(\lambda)
=\Det\!\left(
\lambda I - \sum_{i\in  \mathsf X}  M_i \otimes \bar{M}_i
\right) = 0,
\end{equation}
where $\{ M_i\}_{i\in\mathsf X}$ is a set of Kraus operators for ${\cal M}$ and $\bar{M}_i$ is the operator obtained by taking the entry-wise complex conjugate of $M_i$ with respect to a selected basis of ${\cal H}$.
\begin{enumerate}
\setcounter{enumi}{3}
\item The spectrum of $\cal M$ is invariant under complex conjugation:  if $\lambda  \in  \mathbb C $  is an eigenvalue with eigenvector $A$, then its complex conjugate $\bar \lambda$ is an eigenvalue with eigenvector given by the adjoint operator $A^\dag$, namely 
\begin{equation}
{\cal M}(A^\dag) =\bar{\lambda}A^\dag,   \label{FF11}  
\end{equation} 
with $\bar{\lambda}$ being the complex conjugate of $\lambda$ and $A^\dag$ being the adjoint of $A$. This  property holds not only for quantum channels, but also for all linear maps that are Hermitian-preserving (that is, they send Hermitian operators to Hermitian operators).

\item
 The eigenvalues of a CPTP map ${\cal M}$ are confined in the unit circle on the complex plane. In other words, if there exists a non-zero $A \in {\cal B}({\cal H})$ such
that (\ref{FF}) holds, then 
 we must have $|\lambda| \leq 1$ [this property is indeed a direct consequence of~(iii)].
\end{enumerate} 
 The eigenvalues of ${\cal M}$  which
lie at the boundary of the permitted region, i.e.\ which have unit modulus 
 $|\lambda|=1$, are called \emph{peripheral}.  Of particular interest for us is the unit eigenvalue $\lambda=1$: the associated eigenvectors $A\in  {\cal B}  ({\cal H})$ 
are  called \emph{fixed  points} of ${\cal M}$ to stress the fact that  they are
 left unchanged by the action of the map $\mathcal{M}$:
\begin{enumerate} 
\setcounter{enumi}{5}
\item  Every CPTP map admits \emph{at least one} fixed point state, i.e.\ a solution of (\ref{FF}) for $\lambda=1$, which belongs to the
 set $\mathfrak{S}({\cal H})$ of the density matrices of the system.
\end{enumerate}

As a matter of fact, a generic quantum channel 
possesses  more than just one density matrix that fulfills the requirement~(vi) 
   (for instance unitary transformations admit infinitely many fixed point states).
In the rest of the paper  however we will focus on the special subset of CPTP maps which have exactly  a single
element of $\mathfrak{S}({\cal H})$ that is stable under the transformation:

\begin{defi} \label{def:ergodic}
A CPTP map $\mathcal{M}$ is said to be 
 ergodic if  there exists a unique state $\rho_* \in \mathfrak{S}({\cal H})$ which is  left  unchanged by the channel ${\cal M}$, i.e.\ which solves
 (\ref{FF}) for $\lambda=1$.
 We    introduce the symbol $\mathfrak{C}_E({\cal H})$ to represent
 the subset containing all the ergodic elements of~$\mathfrak{C}({\cal H})$.
 \end{defi}

 As it will be explicitly shown in the next section (see Corollary \ref{lem:unique}), the ergodicity is strong enough to guarantee 
 that $\rho_*$ is not only  the unique $\lambda=1$ solution of (\ref{FF}) on $\mathfrak{S}({\cal H})$ but also  (up to a multiplicative factor) 
 the only solution for the same problem 
  in the larger set of the operators ${\cal B}({\cal H})$. An equivalent (and possibly more intuitive) way to define ergodicity  can be obtained by posing a 
constraint on the  effective \emph{discrete-time} evolution generated by 
the repetitive applications of ${\cal M}${\blue ~\cite{HOLEVObook,oseledets,EVANS,MOROZOVA,DAVIES}}. Specifically, given $\rho  \in \mathfrak{S}({\cal H})$ a  generic input state of the system, 
consider the series 
\begin{equation}\label{average0}
\Sigma_N^{({\cal M})}(\rho) = \frac{1}{N+1} \sum_{n=0}^N {\cal M}^{n}(\rho), 
 \quad \forall \rho\in\mathfrak{S}({\cal H}) ,
\end{equation} 
where ${\cal M}^0={\cal I}$ stands for the identity superoperator, while for $n\geq1$, 
${\cal M}^{n}$ is a CPTP map associated with $n$ recursive applications of ${\cal M}$, i.e.  
 \begin{equation}
 {\cal M}^{n}:= \underbrace{{\cal M} \circ \cdots \circ {\cal M}}_{n\ \mathrm{times}}   \label{CONC}
 \end{equation}
(see for example Refs.~\cite{NJP,Jajte} for an explicit proof of this fact).
Equation (\ref{average0}) describes the average state associated with the first $N+1$ steps of the discrete
trajectory of $\mathfrak{S}({\cal H})$, defined by the  density matrices 
$\rho$, ${\cal M}(\rho)$, ${\cal M}^2(\rho)$, \ldots, ${\cal M}^N(\rho)$, obtained by applying ${\cal M}$ to $\rho$ recursively. 
It then turns out that  ${\cal M}$ is  in $\mathfrak{C}_E({\cal H})$ and has a unique fixed point  state $\rho_*$, if and only if  $\Sigma_N^{({\cal M})}(\rho)$ converges in trace norm to $\rho_*$, i.e.
\begin{equation}
\lim_{N\to\infty} \| \Sigma_N^{({\cal M})}(\rho) - \rho_* \|_1=0 .
\end{equation} 
Accordingly, this implies that for ergodic channels  the average of the  expectation values   $a_n=\Tr[ A {\cal M}^n(\rho)]$ of 
 any bounded operator  $A$, evaluated along the trajectory $\rho$, ${\cal M}(\rho)$,  ${\cal M}^2(\rho)$, \ldots, ${\cal M}^N(\rho)$, converges
 to the fixed point value  $a_* = \Tr[A \rho_*]$, i.e. 
\begin{equation}
\lim_{N\rightarrow \infty}  \frac{1}{N+1} \sum_{n=0}^N a_n = a_*. \label{AAA}
\end{equation}

A proper subset of  $\mathfrak{C}_E({\cal H})$ is constituted by \emph{mixing/relaxing} maps \cite{NJP}:

\begin{defi} \label{def:mixing} 
A quantum channel ${\cal M}$ is said  to be mixing if   $\exists! \rho_* \in \mathfrak{S}({\cal H})$ such that 
 \begin{equation} \label{mixcon}
\lim_{n\rightarrow \infty} \| {\cal M}^n(\rho) - \rho_*\|_1 = 0 ,
 \quad \forall \rho\in\mathfrak{S}({\cal H}) .
\end{equation} 
We call the special state $\rho_*$  the \emph{fixed point state} of ${\cal M}$  and introduce the symbol $\mathfrak{C}_M({\cal H})$ to represent
 the subset containing all the mixing elements of $\mathfrak{C}({\cal H})$.
 \end{defi}

As anticipated  all mixing maps are ergodic (with their  fixed point states provided by the stable density matrices of the mixing channels), i.e.\ $\mathfrak{C}_M({\cal H})\subset \mathfrak{C}_E({\cal H})$.
 Notably however the opposite is not true \cite{NJP}: 
for instance
the qubit channel
\begin{equation}\label{example}
{\cal M}(\rho) = \bra{1}\rho \ket{1}\ket{0}\bra{0} +  \bra{0}\rho\ket{0}\ket{1}\bra{1}
\end{equation} 
is ergodic with fixed point state $\rho_*= (\ket{0}\bra{0} + \ket{1}\bra{1})/2$, but it does not fulfill the  mixing condition (\ref{mixcon})
[in fact,  ${\cal M}^{n}(\ket{0}\bra{0})$  keeps oscillating between $\ket{0}\bra{0}$ and $\ket{1}\bra{1}$]. 
Interestingly in the case of continuous time Markovian evolution, mixing and ergodicity are equivalent (see Section~\ref{semi:SEC}), which means that the above channel cannot be obtained  as the result of a Markovian time evolution  (cf.\ \cite{WOLFCIR}).

The mixing property of a channel ${\cal M}$  can be described in terms of its spectral properties. Indeed a necessary and sufficient condition 
for mixing is the fact that (up to a multiplicative factor) the fixed point state $\rho_*$  of ${\cal M}$ is the unique 
peripheral eigenvector of ${\cal M}$.  More specifically a channel is mixing if and only if it is ergodic \emph{and}  
no solutions exists in ${\cal B}({\cal H})$  for  the eigenvalue equation (\ref{FF}) 
with both $|\lambda| =1$ and $\lambda \neq 1$.

\begin{remark}
Requiring  $\lambda =1$ to be the only peripheral eigenvalue of the channel is not sufficient to enforce the mixing property (or even ergodicity). 
As a counterexample 
consider for instance the case of the identity channel ${\cal I}$. If however $\lambda = 1$ is the only peripheral eigenvalue and has a multiplicity one, then the channel is mixing (see, e.g.\ \cite{WOLF}).
\end{remark}

\section{Characterization of Ergodicity in Terms of Invariant Subspaces}  \label{section3}
 In this section we present a characterization of the ergodicity of a channel $\cal M$ in terms of the subspaces that are left invariant by its action. 
{\red We start  in Section~\ref{sec:fixed} with the discussion on the linear space generated by the fixed points. This introductory subsection collects some  facts  outlined in the seminal works by Davies~\cite{DAVIES}, Morozova and \v{C}encov~\cite{MOROZOVA}, and Evans and H{\o}egh-Krohn~\cite{EVANS}.  Building on these facts, we give a characterization of ergodicity for general channels (without the assumption that the fixed point be a faithful state): a channel is ergodic if and only if it has a \emph{minimal invariant subspace} (Section~\ref{sec:subspaces}).  An equivalent characterization is provided in Section~\ref{sec:ad}, stating that a channel is ergodic if and only if its adjoint map has a maximal invariant subspace. }

\subsection{The linear Space Spanned by Fixed Points}  \label{sec:fixed}
The linear subspace of ${\cal B}  ({\cal H})$ generated by the fixed points of a channel ${\cal M}$ can be shown to be spanned by positive operators.
{\blue  This fact can be easily established for instance by exploiting the following property 
(see e.g.\ Theorem 7.5  of  Ref.~\cite{MOROZOVA} or Proposition 6.8 of Ref.~\cite{WOLF}): }

\begin{lemma}\label{lem:positivefix}
Let $\cal M  \in \mathfrak C  ({\cal H})$ be a quantum channel, and let    $A \in  {\cal B}  ({\cal H})$ be a fixed point of $\cal M$, written as  $ A  =  (X_+  -  X_-)  +  i   (Y_+ -  Y_-)$, where   $X_+, X_-, Y_+, Y_-   \in  {\cal B}_+  ({\cal H}) $  are positive operators such that  $X_+  X_-  {= X_-  X_+}  =  Y_+  Y_- {=  Y_-  Y_+} =  0$.  
Then, also    $X_+, X_-, Y_+, Y_- $ are fixed points of $\cal M$.   
\end{lemma}

\Proof
The proof is provided in the Appendix. 
\qed

\bigskip
\noindent
From this it immediately follows that non-ergodic channels always admit at least two fixed point states that are ``not overlapping":

\begin{cor}\label{cor:orthofix}  
Let $\cal M  \in \mathfrak C ({\cal H})$ be a quantum channel.  If $\cal M$ has two distinct fixed  point states, then $\cal M$ has also two fixed  point states that have  \emph{orthogonal supports}.    
\end{cor}

\Proof
Suppose that two distinct  density matrices  $\rho_0$ and $\rho_1$  are fixed point states  of  ${\cal M}$.  Then, the  traceless operator $\Delta :  =  \rho_0  -  \rho_1  \neq 0$  is also a  fixed point  of the channel.    Denoting by $\Delta_+\ge0  $ and $\Delta_-\ge 0$ the positive and negative parts of $\Delta$, respectively, by Lemma~\ref{lem:positivefix}  we have that  $\Delta_+$  and $\Delta_-$ are both fixed points, and so are the states $\rho_\pm  :  =  \Delta_\pm  /  \Tr[\Delta_\pm]$.  Clearly, $\rho_+$ and $\rho_-$ have orthogonal supports. 
\qed

\bigskip
\noindent
Another   consequence of Lemma~\ref{lem:positivefix} is that all fixed points of an ergodic channel 
must coincide up to a proportionality constant:

\begin{cor}\label{lem:unique}
A CPTP map 
$\mathcal{M}$ is ergodic with fixed point state $\rho_*$, if and only if  all   the solutions in ${\cal B}({\cal H})$ of the eigenvalue equation
\begin{equation}
\mathcal{M}(A) =A\label{eq2}
\end{equation}
can be expressed as 
\begin{equation}
 A = \rho_*\mathop{\emph{Tr}}[A]. \label{eq3}
 \end{equation} 
 \end{cor}

\Proof
If all the solutions of (\ref{eq2}) can be expressed as (\ref{eq3}), then the map $\mathcal{M}$ is clearly ergodic, 
the converse instead follows by contradiction from Lemma~\ref{lem:positivefix}. 
\qed

\bigskip
It is worth noticing that an alternative proof of the Corollary~\ref{lem:unique} can  also be obtained  from Lemma 6 of \cite{NJP},
which states that if  $A$ is a peripheral eigenvector  of a channel ${\cal M}$ then  $|A|:=\sqrt{A^\dag A}$  must be a fixed point of $\cal M$ (see the Appendix for details).

\subsection{Fixed Points and Invariant Subspaces} \label{sec:subspaces}
Here we link the ergodicity property of a channel to the structure of its invariant subspaces, that is, of those subspaces of ${\cal H}$ that are left invariant by the action of the Kraus operators of the channel.  The study of invariant subspaces and their relation to fixed points was previously used in \cite{TICC} as a tool to engineer stable discrete-time quantum dynamics and in \cite{infopres} as a tool to characterize the algebraic structure of the fixed points of a given quantum channel.

\begin{defi} \label{def:invariant}
We say that a subspace $S  \subseteq \cal H$ is \emph{invariant} for a completely positive (not necessarily trace-preserving) map $\cal{M}$     if and only if for every Kraus representation  ${\cal M}  (\rho)  =  \sum_{i \in  \mathsf X} M_i \rho M_i^\dag$ we have $M_i  |\varphi\rangle  \in  S$ for every $|\varphi\rangle \in  S$ and for every $i\in  \mathsf X$. 
\end{defi}

\noindent
It is worth reminding that the condition that $S$  is an invariant subspace under ${\cal M}$ can  be equivalently expressed by the following properties: 
\begin{itemize}
\item[a)]  $ M_i  P  =   P M_i  P$  for every $i  \in\mathsf X$, where $P$ is the projector on $S$;  
 \item[b)] ${\cal M}(P)=P {\cal M}(P) P$;
\item[c)] $ \Supp[{\cal M}  (\rho)]   \subseteq S $ for every state $\rho \in \mathfrak{S}({\cal H})$ with $\Supp (\rho)  \subseteq S$  [here $\Supp (\rho)$ stands for the support of the state $\rho$].
\end{itemize}

\noindent 
The invariant subspaces of a channel are related to its fixed  point states  in the following way:

\begin{lemma}\label{lem:invfix}
If $\rho\in \mathfrak S  ({\cal H})$ is a fixed point  state for $\cal M$, then the support of $\rho$ is an invariant subspace.  
Moreover,  if $S  \subset \cal H$ is an invariant subspace for $\cal M $, then there exists a fixed point state $\rho_S \in  \mathfrak S ({\cal H})$ with $\Supp (\rho_S)  \subseteq S$.  
\end{lemma}

\Proof
From Lemma~\ref{lemma:sup} of the Appendix it follows that  for every unit vector $|\varphi\rangle  \in  \Supp (\rho)$  there exists a positive probability $p>0$ and a state $\sigma$ such that $\rho  =  p |\varphi\rangle\langle  \varphi|   +  (1-p)  \sigma$. 
Furthermore since ${\cal M}(\rho)  =  p {\cal M}(|\varphi\rangle\langle  \varphi|)   +  (1-p)  {\cal M}(\sigma)$, the same Lemma implies that 
we must also have 
\begin{equation}
\Supp [  {\cal M}  (|\varphi\rangle\langle  \varphi| ) ]  \subseteq \Supp [{\cal M} (\rho)]. \label{eq111}
\end{equation}
Consider then the case in which $\rho$ is a fixed point state for ${\cal M}$, i.e.\ ${\cal M}(\rho)=\rho$. 
Equation (\ref{eq111}) then implies that   $\Supp [  {\cal M}  (|\varphi\rangle\langle  \varphi| ) ]  \subseteq   \Supp (\rho) $ for every $\ket{\varphi}\in\Supp(\rho)$, namely that  $\Supp (\rho)$ is an invariant subspace. 
Conversely, let $S$ be an invariant subspace for  $\cal M$. Then the restriction of $\cal M$  to $S$ is a channel in $ \mathfrak C (S) $, and, as such, has a fixed point $ \rho_S \ge 0$ with $\Supp (\rho_S)  \subseteq S$.    
\qed

\bigskip
Using the relation between fixed point states and invariant subspaces  we can obtain a first characterization of ergodicity in terms of the invariant subspaces of the channel:

\begin{theo}[Characterization of ergodicity in terms of invariant subspaces]\label{theo:ergogen}  
For a quantum channel $\cal M  \in \mathfrak C ({\cal H})$, the followings are equivalent:
\begin{enumerate}
\item  $\cal M$ is ergodic;
\item $\cal M$ does not have two invariant subspaces $S_1\not =  \{0\}$ and $S_2 \not =  \{0\}$ such that $S_1  \cap S_2  =  \{0\}$;
\item $\cal M$ has a minimal invariant subspace, that is, a subspace $S \not  = \{ 0\} $ such that  $S  \subseteq S'$ for every invariant subspace $S' \not  = 0$ (remark: the minimal invariant subspace coincides with the support of the fixed point state).
\end{enumerate}       
\end{theo}

\Proof
(i)  $\Rightarrow$  (ii)
If $\cal M$ has two non-intersecting invariant subspaces $S_1$ and $S_2$, then, by Lemma~\ref{lem:invfix} it has two distinct fixed points  $\rho_{S_1}$ and $\rho_{S_2}$, 
respectively.  Hence, $\cal M$ is not ergodic.           
(ii) $\Rightarrow$ (i)
If $\cal M$ is not ergodic, then it has two distinct invariant states. By Corollary~\ref{cor:orthofix}, this implies that $\cal M$ has two orthogonal invariant states, $\rho_+$ and $\rho_-$.  Hence $\Supp(\rho_+)$ and $\Supp (\rho_-)$ are two orthogonal invariant subspaces, and, in particular $\Supp (\rho_+)  \cap \Supp (\rho_-)  =  \{0\}  $.  
(i) $\Rightarrow$ (iii)
If $\cal M$ is ergodic with fixed point $\rho_*$, then $\Supp (\rho_*)$ is a minimal invariant subspace.   Indeed,  for every invariant subspace $S$ there is an invariant state $ \rho_S $ with $\Supp (\rho_S)  \subseteq S$ (Lemma~\ref{lem:invfix}).  Now, since $\cal M$ is ergodic there is only one invariant state, i.e.\ $\rho_S  \equiv \rho_*$.    Hence, $\Supp(\rho_*)  \subseteq S$.       
(iii) $\Rightarrow$ (ii)
If $\cal M$ has two non-intersecting invariant subspaces  $S_1$ and $S_2$, then it cannot have a minimal invariant subspace   $S$, because in that case we should have $S  \subseteq  S_1 \cap S_2  =  \{0\}$.      
\qed

\bigskip
 As an example consider the case of the erasure channel which maps every state into a given selected state $\rho_0$ according to the transformation
\begin{equation}
{\cal M} (A)  =   \rho_0 \Tr[A],  \quad \forall A  \in {\cal B}  ({\cal H }). \label{exera}
\end{equation}
This map is clearly ergodic with the fixed point state being
$\rho_0$. We can then easily verify that in 
agreement with Theorem~\ref{theo:ergogen}  
any invariant subspace $S\neq 0$ of ${\cal M}$  must necessarily contain the support of $\rho_0$. Indeed from condition b)  below Definition~\ref{def:invariant}
it follows that  $S$ is invariant under ${\cal M}$   when given $P$ the
projector on $S$  we have  
\begin{equation}
 \Tr[P]\rho_0 =  \Tr[P]P \rho_0 P,
\end{equation} 
but since $\Tr[P]\neq 0$, this can only be true if  $\Supp(\rho_0)  \subseteq S$.

Theorem \ref{theo:ergogen}  provides a straightforward characterization of ergodicity for random-unitary \emph{qubit} channels $\mathcal M:\mathcal B(\mathbb C^2)\to\mathcal B(\mathbb C^2)$ (the two-dimensionality of the Hilbert space is essential here, as it excludes the case where the unitaries commute on a subspace):

\begin{cor}[Ergodic random-unitary qubit channels] \label{cor:randqubit} 
A random-unitary {\em qubit} channel ${\cal M}   =  \sum_{i\in \sf X} p_i{\cal U}_i$, with ${\cal U}_i  (\rho)  =  U_i \rho U_i^\dag$  being unitary transformations and $p_i>0$ positive probabilities,  is ergodic if and only if within the set
 $\{U_i\}_{i\in \sf X}$ 
there exist at least two elements which  do not commute.
\end{cor}

\Proof
If  all the  unitaries $U_i$ commute, they can be jointly diagonalized, and every joint eigenvector is a fixed point, implying that the channel is not ergodic.  Conversely, if  at least two unitaries of the set  do not commute,  the only invariant subspace $S  \not = 0 $ is $S  =\cal H$.  Hence,  $S$ is clearly minimal and Theorem \ref{theo:ergogen} guarantees that $\cal M$ is ergodic.  
\qed

\begin{remark}
We remind that in the case of qubit channels the set of random-unitaries coincides with the set of unital maps (i.e.\ with the set of CPTP maps which admit the identity operator as fixed point) \cite{UNITALQ}. Corollary~\ref{cor:randqubit}
hence provides  a complete characterization of ergodicity for qubit unital maps. A (partial) generalization of this result to the case of unital maps operating on higher dimensional Hilbert spaces is given in Section~\ref{sec:unital}.
\end{remark}

\subsection{Ergodicity in Terms of the Adjoint Map} \label{sec:ad}
{\blue A fully equivalent description of the ergodicity property of a quantum channel $\cal M$ can be obtained by considering its adjoint map $\cal M^\dag$. As a matter of fact the
original works on ergodicity for quantum stochastic processes  were mostly discussed in this context, the ergodicity of the
original channels been typically discussed as a derived property, see e.g.\ Refs.~\cite{DAVIES,MOROZOVA,EVANS,oseledets}.}

\begin{defi} 
Given a linear map $\cal M:  {\cal B}  ({\cal H})  \to  {\cal B}  ({\cal H})$, the  
   \emph{adjoint} of $\cal M$  is the linear map $\cal M^\dag :  {\cal B}  ({\cal H})  \to  {\cal B}  ({\cal H})$ uniquely defined by the relation  
\begin{equation}\label{adjoint}
\langle   A  ,   {\cal M}  (B)  \rangle  =   \langle  {\cal M}^\dag( A ) ,     B  \rangle,
\end{equation}       
where $ \langle   A  ,   B  \rangle  :  = \mathop{\emph{Tr}}[A^\dag B] $ is the Hilbert-Schmidt product.  
\end{defi}

Going from a quantum channel to its adjoint is the same as going from the Schr\"odinger picture to the Heisenberg picture: if $\cal M$ represents the evolution of the states, then $\cal M^\dag$ represents the evolution of the observables.   It is well known that  $\cal M$ is (completely) positive if and only if $\cal M^\dag$  is (completely) positive and
that $\cal M$ is trace-preserving if and only if $\cal M^\dag$ is identity-preserving, i.e.
\begin{equation}
{\cal M}^\dag (I)  =  I\label{unitalcond}.
\end{equation} 
When $\cal M$ is  completely positive,  a Kraus representation for $\cal M^\dag$ can be obtained by taking the adjoint of the Kraus operators of ${\cal M}$:  if ${\cal M}  (\rho)  =  \sum_{i\in  \mathsf X}  M_i \rho M_i^\dag $, then
\begin{eqnarray}\label{krausdelduale}
{\cal M}^\dag  (A)  =   \sum_{i\in  \mathsf X}   M_i^\dag A  M_i,  \quad \forall A \in  {\cal B}  ({\cal H}) . 
\end{eqnarray} 
As a consequence,  the  spectra of the two maps are identical, i.e.\ they share the same eigenvalues. 

Indeed, (\ref{krausdelduale}) implies that the  characteristic polynomials (\ref{dffd}) of the two maps coincide up to complex conjugation, i.e. 
\begin{equation} \label{dffd1}
{\Poly}^{({\cal M}^\dag)}(\lambda)= 0 \quad \Longleftrightarrow  \quad 
{\Poly}^{({\cal M})}(\bar{\lambda})= 0.
\end{equation}  
Since the spectrum of a completely positive map  is invariant under complex conjugation [cf.\ property (iv) in Section~\ref{sec:1}], this proves that $\cal M $ and $\cal M^\dag$ have the same spectrum.  

In view of the above result it makes sense to extend the definition  of  ergodicity and mixing also for the adjoints of CPTP channels:

\begin{defi} \label{def:newmixing} 
Given a CPTP map ${\cal M}$ we say that its adjoint ${\cal M}^\dag$  is {\em ergodic}
when (up to a proportionality constant) it admits only the operator $I$  as the fixed point
and   {\em mixing} if  furthermore  it does not possess other peripheral eigenvalues.
\end{defi}

\noindent
In other words, 
 $\cal M^\dag$ is defined to be ergodic if $\cal M$ has only trivial \emph{constants of motion}:  an observable $A  \in  {\cal B}  ({\cal H})$ is a constant of motion for the channel $\cal M$ if  for every state $\rho  \in  \mathfrak  S   ({\cal H}) $ we have $  \Tr [A  {\cal M} (\rho)]  =   \Tr  [  A \rho]$, or, equivalently, if ${\cal M}^\dag (A)  =  A$.   Saying that $\cal M^\dag$ is ergodic amounts hence to saying that the only independent constant of motion is the trace of the density matrix.

With the above definitions one can   show that a CPTP map $\cal M$ is ergodic (mixing) if and only if  its adjoint $\cal M^\dag$ is ergodic (mixing):

 \begin{theo}[Ergodicity and constants of motion] \label{th:constant}
A channel ${\cal M}\in  \mathfrak C  (\cal H)$ 
is ergodic
if and only if 
 all the constants of motion are  multiples of the identity [i.e. if and only if 
 adjoint map $\cal M^\dag$ is ergodic.].  
\end{theo}

\Proof 
Consider the quantum channels $\Sigma_N^{({\cal M})}$ defined in (\ref{average0}). As discussed in Section~\ref{sec:1},    if  $\cal M$ is ergodic with fixed point $\rho_*$, 
  then for each bounded operator $A$ and for all density matrices $\rho$ we must have 
\begin{equation}
  \lim_{N\rightarrow \infty}  \langle  A,   \Sigma_N^{({\cal M})}(\rho)   \rangle  =  \langle A, \rho_* \rangle \label{eq12}.
\end{equation}  
By linearity  the adjoint channel of  $\Sigma_N^{({\cal M})}$ is given by $\Sigma_N^{({\cal M}^\dag)}$. Therefore if $A$ is a constant of motion for ${\cal M}$, then it is also 
a constant of motion for $\Sigma_N^{({\cal M})}$,
  i.e.\ $\Sigma_N^{({\cal M}^\dag)}(A) = A$. In this case (\ref{eq12}) can be written as  
  \begin{equation}
\langle A, \rho_* \rangle  =   \lim_{N\rightarrow \infty}  \langle  \Sigma_N^{({\cal M^\dag})}(A) , \rho  \rangle = \langle A, \rho\rangle ,
  \end{equation}    
which, to be true for all $\rho$, implies    $A  = \langle A, \rho_* \rangle  I $. 
Conversely 
suppose 
that all constants of motion of ${\cal M}$ are proportional to the identity.  Assume then by contradiction that  
    $\cal M$ is not ergodic. By Corollary~\ref{cor:orthofix} we know that ${\cal M}$ must have  two orthogonal invariant states $\rho_0$ and $\rho_1$. Take then two orthogonal projectors $P_0\neq 0$ and $P_1\neq 0$ such that $\langle P_i  ,  \rho_j \rangle =  \delta_{ij}$, and define the operators  $P_{i,\infty} :=   \lim_{N\rightarrow \infty} \Sigma_N^{({\cal M^\dag})}(P_i) $ ($i=0,1$).   
    One can easily verify that $P_{0,\infty}$ and $P_{1,\infty}$   are constants of motion of ${\cal M}$ [indeed they verify the identities 
    ${\cal M}^\dag ( P_{i,\infty}) = P_{i,\infty}$]. However 
 since $\rho_0$ and $\rho_1$ are fixed point states of the map,  we also have $\langle P_{i,\infty} , \rho_j   \rangle  = \lim_{N\rightarrow \infty} \langle   P_i , \Sigma_N^{({\cal M})}   (  \rho_j )  \rangle  = \langle   P_i ,     \rho_j   \rangle   = \delta_{ij} $, which is in contradiction to the fact that   $P_{0,\infty}$ and $P_{1,\infty}$  should be  multiples of the identity.    
\qed

\begin{cor} \label{mixing}
A CPTP map ${\cal M}$  is mixing if and only if its adjoint channel ${\cal M}^\dag$ admits the identity operator as unique eigenvector associated with a  peripheral eigenvalue.
\end{cor}

\Proof 
Recall that  ${\cal M}^\dag$  and ${\cal M}$ share the same spectrum.  Having ${\cal M}$ mixing implies that  $\lambda=1$ is the unique
peripheral eigenvalue of ${\cal M}^\dag$. The first implication then follows from Theorem~\ref{th:constant} by recalling that any mixing channel ${\cal M}$ is also ergodic.
Conversely, if ${\cal M}^\dag$ admits the identity operator as unique eigenvector associated with peripheral eigenvalues then by Theorem~\ref{th:constant} ${\cal M}$ is ergodic and no other peripheral
eigenvalues can exist, i.e.\ it is mixing.
\qed

\bigskip
We have already noticed that the spectrum of 
a quantum channel $\cal M$  coincides with that of its adjoint ${\cal M}^\dag$. 
Here we strengthen this result by showing that, at least for peripheral eigenvalues, there is a simple relation which connects the associated eigenvectors:

\begin{lemma}\label{lem:strano}  
Let $\cal M$ be a (not necessarily ergodic) quantum channel with fixed point state $\rho_*$.   If $\omega \in\mathbb C$ with  $|\omega| =1$  is an eigenvalue of $\cal M^\dag$ with eigenvector $ A  \in  {\cal B}  (\cal H) $, then $\bar \omega$ is an eigenvalue of $\cal M$ with eigenvector  $\tilde A  := A\rho_*$.     The converse holds if $\cal M$ has a strictly positive fixed point $\rho_*>  0$:  under this hypothesis,   if $\bar \omega \in\mathbb C$ with  $|\omega| =1$  is an eigenvalue of $\cal M$ with eigenvector $ \tilde A  \in  {\cal B}  (\cal H) $, then $ \omega$ is an eigenvalue of $\cal M^\dag$ with eigenvector  $A  := \tilde A\rho_*^{-1}$.     
\end{lemma}

\Proof 
Suppose that  ${\cal M}^\dag (A)  = \omega A$. Then,   introducing  Kraus operators $\{M_i\}_{i \in  \mathsf X}$ for ${\cal M}$ we have   
\begin{eqnarray}
\Tr [ A^\dag A\rho_*]
&=     \omega   \langle    {\cal M}^\dag (A), A   \rho_* \rangle   
    =    \omega  \sum_{i}   \Tr  [    M_i  A  \rho_*   M_i^\dag  A^\dag   ]         \nonumber\\
   &
   \le \sqrt{\sum_i  \Tr[  M_i A  \rho_*  A^\dag  M^\dag_i  ] \cdot    \sum_j  \Tr[   A  M_j     \rho_*  M_j^\dag   A^\dag   ]   }\nonumber\\
   &
    =   \sqrt{    \Tr [{\cal M}  ( A  \rho_*  A^\dag) ]     \Tr[   A  {\cal M}  (   \rho_*)     A^\dag   ]    } =  \Tr[A^\dag A  \rho_*],
\end{eqnarray}    
where we have used the fact that $\rho_*$ is a fixed point state for ${\cal M}$, the cyclicity of the trace,  and the Cauchy-Schwarz inequality. 
Since the equality can only be obtained  when the Cauchy-Schwarz inequality is saturated,
 we must have  $ \omega  M_i  A  \rho_*^{1/2}  =    A M_i \rho_*^{1/2}$ for all $i$.  Multiplying by ${\rho_*^{1/2}}M_i^\dag$  on the right and summing over $i$, we obtain  
the identity $\omega  {\cal M}  (A \rho_*)  =  \omega \sum_i  M_i  A  \rho_*  M_i^\dag  =     A {\sum_i} M_i  \rho_*   M_i^\dag  =  A  {\cal M}  (\rho_*)  =   A \rho_*$,  which shows that
$\tilde{A} =A \rho_*$ is indeed the eigenoperator of ${\cal M}$ belonging to the eigenvalue $\bar{\omega}$.
Conversely, suppose that $\cal M$ has a strictly positive fixed point $ \rho_* >0$ and that $  {\cal M}( \tilde A )  = \bar \omega \tilde A$.   Defining $A  =   \tilde A  \rho_*^{-1}$ we have   
 \begin{eqnarray}
 \Tr [ A^\dag A\rho_*]  
&=  \omega  \langle   A  ,  {\cal M}  (A  \rho_*)  \rangle    
     =    \omega  \sum_{i}   \Tr  [    M_i  A  \rho_*   M_i^\dag  A^\dag   ]      \nonumber\\
   &
   \le \sqrt{  \sum_i  \Tr[  M_i A  \rho_*  A^\dag  M^\dag_i  ]  \; \sum_j  \Tr[   A  M_j     \rho_*  M_j^\dag   A^\dag   ]}\nonumber\\
   &
    =  \sqrt{    \Tr [{\cal M}  ( A  \rho_*  A^\dag) ]     \Tr[   A  {\cal M}  (   \rho_*)     A^\dag   ]    }=  \Tr[A^\dag A  \rho_*].
\end{eqnarray}
Again, to attain the equality in the Cauchy-Schwarz inequality, we must have  $ \omega  M_i  A  \rho_*^{1/2}  =    A M_i \rho_*^{1/2}$ for all $i$, or equivalently,  $ \omega  M_i  A   =    A M_i $ for all $i$. Hence, we have $ {\cal M}^\dag (A)  =   \sum_i  M_i^\dag A  M_i  =   \omega {\sum_i} M_i^\dag M_i A  =\omega {\cal M}^\dag (I)    A   =  \omega A$.  
\qed

\bigskip
\noindent
 It is worth noticing that, for   $\omega=1$,  Lemma~{\ref{lem:strano}} is trivially verified by the eigenoperator $I$ of ${\cal M}^\dag$.

The ergodicity of a CPTP map ${\cal M}$ can also be characterized in terms of the invariant subspaces of $\cal M^\dag$.   
For this purpose, observe that the invariant subspaces of a positive map $\cal M$ are related to the invariant subspaces of its adjoint $\cal M^\dag$  in the following way:

\begin{lemma}\label{lem:perpdag}
Let $\cal M$ be a positive map.  Then $S  \subseteq \cal H$ is an invariant subspace for $\cal M$ if and only if $S^\perp :  =  \{|\varphi \rangle \in  {\cal H}\,|\,\langle \varphi  |  \psi\rangle =  0 , \ \forall | \psi\rangle \in S  \}$ is an invariant subspace for $\cal M^\dag$.  
\end{lemma}

\Proof
Let us denote by $P$ and $P^\perp$  the projectors on $S$ and $S^\perp$, respectively.   For a   completely positive map  $\cal M$ we have that $S$ is invariant if and only if ${\cal M} (P)  =    P {\cal M} (P)  P $ [see the equivalent condition b) below Definition~\ref{def:invariant}]. 
If this constraint  is satisfied, then we have 
 \begin{eqnarray}
 \langle {\cal M}^\dag ( P^\perp),   P \rangle  
=   \langle P^\perp, {\cal M} (P)  \rangle 
 & =  \langle P^\perp,  P   {\cal M} (P)    P\rangle 
 \nonumber\\
 & =  \langle  P P^\perp  P, {\cal M} (P)  \rangle   =  0,  
 \end{eqnarray}  
  which implies  $P{\cal M}^\dag ( P^\perp)  P  =0$.  By positivity of ${\cal M}^\dag ( P^\perp)$, we necessarily have $ {\cal M}^\dag ( P^\perp)   =  P^\perp {\cal M}^\dag ( P^\perp)  P^\perp $.   Hence, $S^\perp$ is an invariant subspace for $\cal M^\dag$. Repeating the same argument, we have that if $S^\perp$ is invariant for $\cal M^\dag $, then $S$ is invariant for $\cal M$. 
\qed

\bigskip
Using this fact, the characterization of Theorem~\ref{theo:ergogen} becomes

\begin{theo}[Characterization of ergodicity in terms of invariant subspace of the adjoint map]\label{Thm:3}
For a quantum channel $\cal M  \in \mathfrak C ({\cal H})$, the followings are equivalent:
\begin{enumerate}
\item  $\cal M$ is ergodic;
\item $\cal M^\dag$ does not have two invariant subspaces $S_1 \not =  \cal H $ and $S_2  \not =  \cal H$ such that  $\Span (S_1  \cup S_2)  =  \cal H$;
\item $\cal M^\dag$ has a maximal invariant subspace, that is, a subspace $S \not  = \cal H $ such that  $S  {{}\supseteq{}} S'$ for every invariant subspace $S' \not  = \cal H$ (remark: the
maximal invariant subspace coincides with the kernel of the fixed point of $\cal M$).
\end{enumerate}       
\end{theo}

Note that in the above Theorem~\ref{Thm:3} the maximal invariant subspace $S$  can consist only of the zero vector,  if the minimal invariant subspace of $\cal M$ is the whole Hilbert space.

{\blue 
\section{Ergodicity and Mixing under Randomization}\label{convex:SEC}}
It is known that when we prepare a (nontrivial)  convex combination of a
  mixing channel with a  generic (not necessarily mixing) quantum channel, the resulting transformation  
  is  also mixing \cite{MEDIATED,MEDIATED2,MEDI1}. {\blue This implies that mixing channels are stable under randomization, or in a more formal 
  language, that 
  $\mathfrak{C}_M({\cal H})$
  constitutes  a convex subset, which is dense in  $\mathfrak{C}({\cal H})$. 
   Aim of this section is to extend this analysis showing that  
    the same property holds for the larger set  of 
   ergodic channels $\mathfrak{C}_E({\cal H})$  (notice that this fact cannot be established by simple geometric 
    arguments based upon the fact that $\mathfrak{C}_E({\cal H})$ includes $\mathfrak{C}_M({\cal H})$). We also prove 
     a rather counterintuitive fact, namely that the mere action of randomizing an ergodic (not necessarily mixing) channel with the identity mapping
     is capable of introducing mixing into the system. 
}

\medskip
\begin{theo}[Stability of ergodicity under randomization]  \label{theo:ergoconv} 
Let $\mathcal{M}$ be an ergodic CPTP map  and $\mathcal{M}'$ an arbitrary (not necessarily ergodic) element of   $\mathfrak{C}({\cal H})$.
Then, for all $p\in (0,1]$,  the CPTP map 
\begin{equation} \label{convergo}
\mathcal{M}_p:= p\mathcal{M}+(1-p)\mathcal{M}',
\end{equation}
is also ergodic.  Moreover, denoting by  $\rho_*$ and $\rho_{*,p}$ the fixed point states of $\cal M$ and ${\cal M}_p$, respectively, we have that 
\begin{equation}\label{mixturefix}
\rho_{*,p}  =   \pi_p  \rho_*  +   ( 1-\pi_p )  \sigma_p   
\end{equation}
for some probability $\pi_p  \in  ( 0,1]$ and for some state $\sigma_p\in\mathfrak S  (\cal H)$.
\end{theo}

\Proof
Let $\rho\in\mathfrak S (\cal H)$ be a fixed point for ${\cal M}_p$, so that we have   $  p  {\cal M}  (\rho)  +  (1-p)  {\cal M}'  (\rho) =  \rho $.   
By Lemma~\ref{lemma:sup} of the Appendix this relation requires $\Supp  [  {\cal M}  (\rho)]    \subseteq \Supp (\rho)$, that is, $\Supp (\rho)$ is an invariant  subspace for $\cal M$.  Since $\cal M$ is ergodic, by Theorem \ref{theo:ergogen} it has a minimal invariant subspace $S$, and hence $S \subseteq \Supp (\rho)$.  Now, by Theorem \ref{theo:ergogen}, ${\cal M}_p$ must be ergodic, because it cannot have two orthogonal invariant subspaces (every invariant subspace of ${\cal M}_p$  must contain  $S$).   Moreover, recalling that the minimal invariant subspace of $\cal M$  is $S  =  \Supp (\rho_*)$, we obtain the relation  $\Supp (\rho_*)  \subseteq \Supp (\rho_{*,p})$, which implies (\ref{mixturefix}) via Lemma~\ref{lemma:sup}.
\qed

\bigskip
{\blue We also present an alternative proof of the first part of Theorem~\ref{theo:ergoconv}, namely that 
convex combinations of the form (\ref{convergo}) with $\mathcal{M}$ ergodic and $\mathcal{M}'$ generic CPTP maps are also ergodic. 
Differently from the above proof, this does not exploit the connection between ergodicity and invariant subspaces.
Instead it makes use of the following necessary and sufficient condition for the ergodicity:

\begin{cor} \label{lem:nec} A CPTP map $\mathcal{M}$ is ergodic,
if and only if 
$\forall\rho\neq\rho'$ states and $\forall N,\lambda$ positive constants,  we have 
\begin{equation}
\left\|
\sum_{n=0}^N\lambda^n\mathcal{M}^n(\rho-\rho')
\right\|_1 <  
f_N(\lambda)
\|\rho-\rho'\|_1,
\label{eqn:Eq12}
\end{equation}
where  
\begin{equation}
f_N(\lambda)=\sum_{n=0}^{N}\lambda^n  =  
\left\{
\begin{array}{ll}
\medskip
 N+1  &  (\lambda  = 1),   \\
\displaystyle
 \frac{1-\lambda^{N+1}}{  1-\lambda}   & (\lambda \not = 1)  .
\end{array}
\right.
\end{equation} 
\end{cor}

\Proof
If (\ref{eqn:Eq12}) holds then $\mathcal{M}$ is clearly ergodic. Indeed if this is not the case 
${\cal M}$ must have at least two  different fixed points $\rho_*\neq\rho_*'$, which verify the identity  
\begin{equation}
 \left\|
\sum_{n=0}^N\lambda^n\mathcal{M}^n(\rho_*-\rho_*')
\right\|_1
= f_N(\lambda)
\|\rho_*-\rho_*'\|_1,\quad
\forall N>0,\ \forall\lambda>0,
\end{equation}
 contradicting  the assumption. Conversely, assume ${\cal M}\in\mathfrak{C}_E({\cal H})$. For any pair of states $\rho\neq\rho'$ and for any $N>0$ and $\lambda>0$, we have
\begin{eqnarray} 
 \left\|
\sum_{n=0}^N\lambda^n\mathcal{M}^n(\rho-\rho')
\right\|_1
&\le
\sum_{n=0}^N\lambda^n\|\mathcal{M}^n(\rho-\rho')\|_1
\nonumber  \\
&\le
\sum_{n=0}^N\lambda^n\|\rho-\rho'\|_1
=
f_N(\lambda)  \|\rho-\rho'\|_1,
\label{eqn:Eq14}
\end{eqnarray} 
where the first inequality follows from the triangular inequality of the trace distance, while the second from the nonexpansiveness of the CPTP maps ${\cal M}^n$. 
To conclude the proof we need to show that this upper bound cannot be saturated. The first  inequality of (\ref{eqn:Eq14}) can be turned into an identity, if and only if  the operators $\mathcal{M}^n(\rho-\rho')$ ($n=0,\ldots,N$) are all ``parallel,'' i.e.\ $\mathcal{M}^n(\rho-\rho')
=\mu_n(\rho-\rho')$ with $0\le\mu_n\le1$ ($n=0,\ldots,N$).
This implies $\mu_n=\mu_1^n$ ($n=0,\ldots,N$), so that the second inequality in (\ref{eqn:Eq14})  reduces to
\begin{equation}
\sum_{n=0}^N(\lambda\mu_1)^n\|\rho-\rho'\|_1 \le
\sum_{n=0}^N\lambda^n\|\rho-\rho'\|_1 , 
\end{equation}
showing that it  can be transformed into an equality if and only if $\mu_1=1$.
Replacing this into the  parallelism constraint
 we can conclude that the upper bound of (\ref{eqn:Eq14})   can be  saturated if and only if 
$\mathcal{M}(\rho-\rho')=\rho-\rho'$.
However, since ${\cal M}$ is ergodic, from Corollary~\ref{lem:unique} we must have 
$\rho-\rho'=\rho_* \Tr[ \rho-\rho'] = 0$, which contradicts the assumption $\rho\neq \rho'$. 
Therefore, the upper bound of (\ref{eqn:Eq14}), i.e.\ of (\ref{eqn:Eq12}), cannot be saturated for any pair of states $\rho\neq\rho'$.
\qed

\bigskip
We are now in a position to present our alternative proof  that ${\cal M}_p$ of (\ref{convergo})  is ergodic.

\medskip
\Proofa{Proof of Theorem~\ref{theo:ergoconv}}
Note that for every integer $n$ one can write 
$\mathcal{M}_p^n= p ^n\mathcal{M}^n+(1-p^n)\mathcal{S}_n$,
where $\mathcal{S}_n$ is a CPTP map.
Therefore, for $N>0$ and $\lambda>0$ we can invoke the triangular inequality of the trace norm to state that  
\begin{eqnarray}
\left\|
\sum_{n=0}^N\lambda^n {\mathcal{M}^n_p}(\rho-\rho')
\right\|_1
\leq{}&  \left\|
\sum_{n=0}^N\lambda^n p^n\mathcal{M}^n(\rho-\rho')
\right\|_1
\nonumber\\
&{}+\left\|
\sum_{n=0}^N\lambda^n(1-p^n)\mathcal{S}_n(\rho-\rho')
\right\|_1. \label{final}
\end{eqnarray}
Since ${\cal M}$ is ergodic, we can use Corollary~\ref{lem:nec} to bound the first term  as follows
\begin{equation}
\left\|
\sum_{n=0}^N\lambda^n p^n
\mathcal{M}^n(\rho-\rho')
\right\|_1
<
f_N(\lambda p )  
\|\rho-\rho'\|_1
\label{eqn:ThBound} .
\end{equation}
On the contrary using again the triangular inequality and the nonexpansiveness of ${\cal S}_n$ we have 
\begin{eqnarray}
\left\|
\sum_{n=0}^N\lambda^n(1-p^n)\mathcal{S}_n(\rho-\rho')
\right\|_1
&\le
\sum_{n=0}^N\lambda^n(1-p^n)\|\rho-\rho'\|_1
\nonumber\\
& =
[ f_N(\lambda) - f_N(\lambda p) ]\|\rho-\rho'\|_1.
\end{eqnarray}
Substituting this  and (\ref{eqn:ThBound}) in (\ref{final})  we arrive at
\begin{equation}
\left\|
\sum_{n=0}^N\lambda^n {\mathcal{M}_p}^n(\rho-\rho')
\right\|_1
<
f_N(\lambda)  
\|\rho-\rho'\|_1,
\label{trick}
\end{equation}
which according to Corollary~\ref{lem:nec} is sufficient to claim the ergodicity of ${\cal M}_p$.  
\qed 
}

\subsection{From Ergodicity to Mixing}  \label{sec5}
Once established that a channel $\cal M$ is ergodic, one may further ask whether it is also mixing.  
To answer to this question we have to study the peripheral eigenvalues of $\cal M$ 
and to see whether or not $\omega=1$  is the only peripheral eigenvalue.  A useful  observation in this direction is the following results which
 provide a refinement of Lemma 6 of \cite{NJP}:

\begin{lemma}\label{lemma:nuovo}
Let  $\mathcal{M}  \in  \mathfrak  C (\cal H)$ be a (not necessarily trace-preserving) CP map with  Kraus operators $\{ M_i\}_{i \in\mathsf X}$. Then an operator $A\neq 0$ with 
polar  decomposition $A  =  U |A|$ is 
an eigenvector of ${\cal M}$ belonging to the eigenvalue $\omega \in\mathbb C$ if $|A|$ is a fixed point of ${\cal M}$ and if the following condition holds:
\begin{equation}\label{algerho1}
M_i  U |A|  =  \omega   U  M_i  |A| , \quad \forall i\in\mathsf X  .
\end{equation}      
\end{lemma}

\Proof
The thesis immediately follows by observing that if (\ref{algerho1}) holds then we have 
\begin{eqnarray}
{\cal M} (A)
= \sum_{i\in\mathsf X} M_i  U |A| M_i^\dag  
&=  \omega   U \sum_i  M_i  |A|  M_i^\dag    
\nonumber \\
&
=\omega U {\cal M}(|A|) = \omega U |A| = \omega A,
\label{algerho111}
\end{eqnarray}     
where in the second to last passage we have used the fact that $|A|$ is a fixed point of ${\cal M}$. 
\qed

\bigskip
Quite interestingly, when considering peripheral eigenvalues of a trace-preserving CP map  the sufficient condition of Lemma~\ref{lemma:nuovo} can be transformed into a necessary one:

\begin{theo}[Peripheral eigenvectors of quantum channels]\label{theo:ergospec}
Let  $\mathcal{M}  \in  \mathfrak  C (\cal H)$ be a CPTP map,  $\omega  \in\mathbb C$ be a complex number on the unit circle ($|\omega| = 1$), and $A\in  {\cal B}  ({\cal H})$ be an operator with polar decomposition $  A  =  U |A|$.   
The followings are equivalent:
\begin{enumerate}
\item $A$  is an eigenvector of $\mathcal M$ with eigenvalue $\omega$;
\item $|A|$ is a fixed point of $\cal M$ and
\begin{equation}\label{algerho}
M_i  U |A|  =  \omega   U  M_i  |A|,\quad \forall i\in\mathsf X  ,   
\end{equation}      
where ${\cal M} (\rho)  =  \sum_{i\in\mathsf X}  M_i \rho M_i^\dag $ is an arbitrary Kraus decomposition of $\cal M$. 
\end{enumerate}
In particular, if $\cal M$ is ergodic, then   (up to a multiplicative constant) $A$ must be of the form $A  =  U  \rho_*$, where $U$ is unitary and $\rho_*$ is the unique fixed point state of $\cal M$.
\end{theo}

\Proof 
The eigenvalue condition ${\cal M}(A)=\omega A$  can be reformulated as $ |A|   =  \bar \omega  \sum_i  (U^\dag M_i  U)  |A|   M_i^\dag $.  Hence, we have  
\begin{eqnarray}
\Tr[  |A|]  
&= \bar \omega  \sum_i    \Tr[(U^\dag M_i  U)  |A|   M_i^\dag ]\nonumber\\
  &\quad  \le    \sqrt{  \sum_i    \Tr[(U^\dag M_i  U)  |A| (U^\dag M^\dag_i  U) ]\cdot \sum_j    \Tr[ M_j   |A|  M^\dag_j ]}    \nonumber\\
  & \quad =   \sqrt{   \Tr[{\cal M}   (  U^\dag |A |   U   )]     \Tr[{\cal M}   (  |A |   )]  }
  =    \Tr[ |A |].  
\end{eqnarray}
Since the Cauchy-Schwarz inequality is saturated, we necessarily have $ |A|^{1/2}  M_i^\dag   =  \omega  |A|^{1/2}  U^\dag  M_i^\dag   U $  for every $i$, or equivalently,  $ M_i    |A|     = \bar \omega  U^\dag M_i U |A|   $ for every $i$ [in turn, this is equivalent to (\ref{algerho})].      Hence, we have 
\begin{eqnarray}
{\cal M} (|A|)  = \sum_i  M_i  |A|  M_i^\dag
&  =  \bar \omega \sum_i  U^\dag  M_i  U  |A|   M_i^\dag
\nonumber\\
&=     \bar \omega U^\dag {\cal M}  (A)  =  U^\dag  A  =  |A|  ,
\end{eqnarray} 
namely $|A|$  is a fixed point of $\cal M$.   
 The converse is just the statement of Lemma~\ref{lemma:nuovo}.
 Finally, when $\cal M$ is ergodic, the fixed point $|A|$ must be proportional to  $\rho_*$, by Corollary~\ref{lem:unique}.
\qed

\begin{remark}
If ${\cal M}$ is ergodic and $\omega=1$ then Corollary~\ref{lem:unique} implies that the unitary $U$ in Theorem \ref{theo:ergospec} must be the identity operator.
\end{remark}

\begin{remark}
Note that Theorem \ref{theo:ergospec} contains the statement that $|A|$  is a fixed point of $\cal M$ whenever $A$ is an eigenvector of $\cal M$ for some peripheral eigenvalue  $\omega$ with $|\omega| = 1$.  This is the statement of Lemma  6 of \cite{NJP}, of which Theorem \ref{theo:ergospec} provides an alternative derivation.   
\end{remark}

Theorem~\ref{theo:ergospec}
 implies a rather counterintuitive fact: whenever we mix an ergodic channel with the identity channel we necessarily obtain a mixing channel!

\begin{cor}[The mixture of an ergodic channel and the identity is mixing] 
Let ${\cal M}  \in  \mathfrak C( { \cal H})$ be an ergodic channel.  If the linear span $\Span_{\mathbb C} \{ M_i\}_{i \in  \mathsf X}$ contains the identity, then ${\cal M}$ is mixing.  
In particular, if $\cal M$ is an ergodic  channel and ${\cal I}$ is the identity channel, then
\begin{equation}
{\cal M}' =  p {\cal M}  +  (1-p)  \cal I  \label{convexid}
\end{equation}
is mixing for every $p \in (0,1)$.
\end{cor}

\Proof
Let $\omega$ be a peripheral eigenvalue of ${\cal M}$. Since ${\cal M}$ is ergodic Theorem~\ref{theo:ergospec} implies  $M_i  U  \rho_*  =  \omega U M_i  \rho_*$ for all $i\in\mathsf X$, with $U\rho_*$ being the
associated eigenoperator. 
By linearity this equation yields $M U \rho_* =   \omega U M\rho_*$, $\forall M  \in \Span_{\mathbb C} \{ M_i\}_{i \in  \mathsf X}$. Now, if the linear span contains the identity we have $  U\rho_*  =  \omega U\rho_*$, and, therefore $\omega = 1$.  Hence, $\mu = 1$ is the only peripheral eigenvalue of $\cal M$. Since $\cal M$ is ergodic, this implies that $\cal M$ must be mixing.   In particular, the channel (\ref{convexid})
is ergodic by Theorem~\ref{theo:ergoconv}, and obviously  contains the identity among its Kraus operators.   
\qed

\bigskip
 An alternative and instructive  proof of the fact that 
 convex combinations of the identity with an ergodic channel ${\cal M}$  produce mixing maps can  be derived as follows:

\begin{prop}[Spectral properties of  mixtures with  the identity channel] 
Let ${\cal M}  \in  \mathfrak C( { \cal H})$ be a CPTP  channel and $p \in (0,1)$ be a probability. The map  ${\cal M}' =  p {\cal M}  +  (1-p)  \cal I$ admits $\omega=1$ as its unique peripheral eigenvalue.
Furthermore if  ${\cal M}$ is ergodic then ${\cal M}'$ is mixing. 
\end{prop}

\Proof   
If $\omega=e^{i \beta}$ is a generic peripheral eigenvalue of ${\cal M}'$ then by construction  its associated eigenoperator $A$  must satisfy the relation 
 \begin{equation}
 {\cal M}(A) = \frac{\omega - (1-p)}{p}A, 
 \end{equation} 
 where we have used the fact that $p\neq 0$. This in particular implies that $A$ must also be an eigenoperator of ${\cal M}$ belonging to an eigenvalue $\lambda= [\omega - (1-p)]/p$.
 Since ${\cal M}$ is CPTP, however, we must have $|\lambda|\le 1$, i.e. 
 \begin{equation}
 \sqrt{1 + {2}\frac{1-p}{p^2} (1- \cos \beta )} \le 1,
 \end{equation}  
 which, since $p\neq 1$,  can only be true if $\beta =0$, i.e.\ $\omega=1$  and $\lambda=1$.
Moreover in case  ${\cal M}$ is ergodic we can invoke Corollary~\ref{lem:unique} to claim that $A$ must be proportional to its fixed point state $\rho_*$. It then follows that (up to a proportionality constant) the only peripheral eigenvector of ${\cal M}'$ 
  coincides with $\rho_*$: the channel is hence mixing, with its stable point being~$\rho_*$.       
\qed

\bigskip
We conclude the section by observing that Theorem~\ref{theo:ergospec}  allows one to give an alternative proof of the well-known fact about the stability of the property of mixing under convex randomizations:

\begin{theo}[Stability of mixing under randomization]  
If the channel $\cal M\in \mathfrak C (\cal H) $ is mixing, then the channel ${\cal M}_p  :=  p  {\cal M}  +  (1-p)   {\cal M}'$ is mixing for every $p  \in  (0,1]$ and for every channel ${\cal M}'  \in   \mathfrak C (\cal H)$.  
\end{theo}

\Proof  
Suppose that  ${\cal M} (\rho)  =  \sum_{i\in\mathsf X}  M_i \rho M_i^\dag$ is mixing and let $\rho_*$ be its fixed point.   Then, for every operator $U$,  the equation $M_i  U \rho_*  =  \omega   U  M_i\rho_*$, $\forall i\in\mathsf X$  implies $\omega =1$ (otherwise $\cal M$  would have a peripheral eigenvalue $\omega \not =  1$).  
Moreover, since $\cal M$ is mixing,  we know from Theorem~\ref{theo:ergoconv} that ${\cal M}_p$ is ergodic for every $p \in  (0,1]$. Hence, from Theorem~\ref{theo:ergospec} it follows that any peripheral eigenvector of ${\cal M}_p$ is of the form $ A  = U  \rho_{p,*}$, where $\rho_{p,*}$ is the fixed point of ${\cal M}_p$  and $U$ is a unitary satisfying $M_{p,i}  U \rho_{p,*}  =  \omega   U  M_{p,i}\rho_{p,*}$, $\forall i\in\mathsf X_p$, with $\{M_{p,i}\}_{i  \in \mathsf X_p}$ being the Kraus operators of ${\cal M}_p$.  Multiplying by the inverse of $\rho_{p,*}$ on its support, we then obtain  
\begin{equation}
M_{p,i}  U  Q_p   =  \omega   U  M_{p,i}  Q_p  ,  \quad \forall i\in\mathsf X_p,
\end{equation}
where $Q_p$ is the projector on the support of $\rho_{p,*}$.   Now, since by Theorem~\ref{theo:ergoconv} the support of $\rho_*$ is contained in the support of $\rho_{p,*}$, we have $Q_p  \rho_*  =  \rho_*$, and therefore,
\begin{equation}
M_{p,i}  U  \rho_*  =  \omega   U  M_{p,i}  \rho_*  ,  \quad \forall i\in\mathsf X_p.
\end{equation}
Since there is a Kraus form for  ${\cal M}_p$ that includes all the  Kraus operators of $\cal M$, this implies in particular $M_{i}  U  \rho_*  =  \omega   U  M_{i}  \rho_*$, $\forall i\in\mathsf X$.  From the fact that $\cal M$  is mixing we conclude that $\omega=1$. Hence,  ${\cal M}_p$ is mixing.   
\qed

\section{Ergodicity and Mixing for Channels with Faithful Fixed Point}    \label{faithfulsec}
{\blue The  characterization } of ergodic channels, provided by Theorem~\ref{theo:ergogen},  becomes more specific in the case of channels with a \emph{faithful} fixed point,  namely a fixed point state $\rho_*$  with $\Supp (\rho_*)  =  \cal H$, or, equivalently $\rho_*  >  0$.  Ergodic channels with faithful fixed point are also known as \emph{irreducible quantum channels}{\blue ~\cite{DAVIES,WOLF} (in the same references, mixing channels with faithful fixed point state are referred to as \emph{primitive}).       
As already mentioned in the introduction, the majority of the results obtained in the field were explicitly derived for this specific maps (see e.g.\ Refs.~\cite{oseledets,wielandt}).}

\begin{theo}[Ergodicity and proper invariant subspaces]\label{theo:ergoinv}
If a channel ${\cal M}  \in  \mathfrak {C}(\cal{H})$  has a faithful fixed point state $\rho_* >0 $, then the followings are equivalent:  
\begin{enumerate}
\item $\cal M$ is ergodic;
\item $ \cal M$ has no proper invariant subspace;
\item $ \cal M^\dag$ has no proper invariant subspace.   
\end{enumerate}
\end{theo}

\Proof  
(i) $\Longleftrightarrow$ (ii)    
Suppose that $\cal M$ has a proper invariant subspace $S\subset \cal H$.  Then, by Lemma~\ref{lem:invfix}, $\cal M$ must have a fixed point $\rho_S$ with $\Supp (\rho_S)  \subseteq S$.   Hence, $\cal M$ has two distinct fixed points $\rho_S$ and $\rho_*$, that is, $\cal M$ is not ergodic.  Conversely, if $\cal M$ is not ergodic, then it has two orthogonal invariant subspaces (Theorem~\ref{theo:ergogen}), which, by definition, are proper subspaces of $\cal H$.   [It is worth stressing that  the implication (ii) $\Rightarrow$ (i)  does not require $\rho_*$ to be of full rank]. (ii) $\Longleftrightarrow$ (iii)   Immediate from Lemma~\ref{lem:perpdag}.    
\qed

\bigskip
Not having a proper invariant subspace is an important algebraic property, which is equivalent to the irreducibility of the matrix algebra $\cal A_{\cal M}$ generated by the  Kraus operators $\{ M_i\}$
~\cite{HOLEVObook}. $\cal A_{\cal M}$ consists of all possible products and linear combinations of products of the Kraus operators. We summarize here this algebraic property presenting it in the form of the following theorem

\begin{theo}[Ergodicity and irreducibility of matrix algebras] 
For a completely positive (not necessarily trace-preserving) map  $\cal M$,  the followings are equivalent:  
\begin{enumerate}
\item $ \cal M$ has no proper invariant subspace   $S \subset \cal H$;
\item the matrix algebra $\cal A_{\cal M}$ generated by the Kraus operators $\{  M_i\}_{i \in  \mathsf X}$ is irreducible;
\item for every vector $\varphi \in  \cal H$, the set of vectors $\{  M_{i_1}   M_{i_2}  \cdots  M_{i_N}  |\varphi \rangle\,|\,(i_1, i_2, \dots, i_N)   \in  \mathsf X^{\times N} , N  \in \mathbb N  \}$ spans the whole Hilbert space.
\end{enumerate}\label{theo:irreducibility}
\end{theo}

\Proof
(i) $\Longleftrightarrow$ (ii)   By definition,  $\cal A_{\cal M}$ is reducible if and only if it has a proper invariant subspace $S \subset \cal H$.   
(i) $\Longleftrightarrow$ (iii)     The span of the vectors   $\{  M_{i_1}   M_{i_2}  \cdots  M_{i_N}  |\varphi \rangle\,|\,(i_1, i_2, \dots, i_N)   \in  \mathsf X^{\times N} ,  N \in  \mathbb N   \}$ is an invariant subspace  $S$. If $\cal M$  has no proper invariant subspace, then $S  =  \cal H$.      Conversely, if $\cal M$ has a proper invariant subspace $S \subset \cal H$, then  given $|\varphi\rangle\in S$ the  vectors   $\{  M_{i_1}   M_{i_2}  \cdots  M_{i_N}  |\varphi \rangle\,|\, (i_1, i_2, \dots, i_N)   \in  \mathsf X^{\times N}   \}$  can only span a subset of $S$. 
\qed

\begin{remark} 
Note that Theorem~\ref{theo:irreducibility} applies both to $\cal M $  and to $\cal M^\dag$: the two algebras  $\cal A_{\cal M}$   and  $\cal A_{\cal M^\dag}$   must be both irreducible.  
\end{remark} 

\begin{remark} 
Condition (iii) in Theorem~\ref{theo:irreducibility} states that  an arbitrary input state $|\varphi\rangle  \in  \cal H$  evolving under the discrete-time dynamics $\cal M$ will generate a stochastic trajectory  of pure states $M_{i_1} M_{i_2} · · M_{i_N} |\varphi\rangle$ whose span cover  the whole Hilbert space.
Such a property is similar in spirit to the classical property that a generic trajectory of an ergodic dynamical system is dense in the state space.  Note however that in the classical case one can have ergodicity for reversible Hamiltonian dynamics, while in the quantum case one can have ergodicity only for quantum channels representing irreversible evolutions
 \end{remark}

\begin{remark}
In fact,  condition (iii) in Theorem~\ref{theo:irreducibility} can be refined by showing that only a finite number $N$ of iterations of the channel are enough for the trajectory of a arbitrary pure state to span the whole Hilbert space. The number $N$ is upper bounded by the quantum Wielandt's inequality \cite{wielandt}, which in our notation reads  $N  \le  d^2(  d^2  -|\mathsf X|  -1)$, where $|\mathsf X|$ is the number of Kraus operators.    
\end{remark}

Further information about the structure of an ergodic map with faithful fixed point $\cal M$ can be extracted from the analysis of its peripheral  eigenvalues and eigenvectors.  This analysis is the subject of a quantum generalization of the Perron-Frobenius theory of classical Markov chains
{\blue -- see e.g.\ Refs. \cite{matrixprod,WOLF2,sarymsakov,oseledets,ALABERIO}.}
The following theorem summarizes some of these results and takes advantage of the complete positivity of quantum channels to give a convenient condition for the peripheral eigenvectors in terms of the Kraus operators:

\begin{theo}[Peripheral eigenvectors of quantum channels with faithful fixed point]\label{theo:perronfrob}
If $\cal M \in  \mathfrak C   ({\cal H})$ is an ergodic channel with faithful fixed point state $\rho_*>0$,   then 
\begin{enumerate} 
\item   an operator $A_\omega \in  \mathcal B (\mathcal H)$  is an eigenvector of $\mathcal M$ with eigenvalue $\omega$ if and only if  (up to a proportionality constant)  $A_\omega  =   U_\omega  \rho_*$, where $U_\omega$ is a unitary operator satisfying 
\begin{equation}\label{multiplicative}
M_i  U_\omega   =   \omega  U_\omega  M_i,  \quad \forall  i \in  \mathsf X,        
\end{equation}    ${\cal M} (\rho)  =  \sum_{i\in  \mathsf X}  M_i  \rho M_i^\dag$ being an arbitrary Kraus decomposition of $\cal M$;
 \item  the peripheral eigenvalues of $\cal M$ are roots of the unit and form a finite cyclic group $\mathsf F=\{  e^{2\pi  i l/L}\,|\, l  =  0, \dots, L-1\}$ with $L  \le d^2$;
 \item every peripheral eigenvalue is non-degenerate;
 \item the unitaries $\{U_\omega\,|\,  \omega \in  \mathsf F\}$ form a unitary representation of the cyclic group $\mathsf F$;
 \item if $\omega$ is a peripheral eigenvalue of $\cal M^\dag$, then the corresponding eigenvector is (proportional to) a unitary;
\item $\omega $ is a peripheral eigenvalue of $\cal M$ with eigenvector $U_\omega \rho_*$  if and only if $\bar \omega$  is a peripheral eigenvalue of $\cal M^\dag$ with eigenvector $U_\omega$. 
\end{enumerate}
\end{theo}

\Proof  
(i) The key of the entire derivation is the identity (\ref{multiplicative}): we hence start proving it.  
  From Theorem~\ref{theo:ergospec}, we know that $A_\omega$ is an eigenvector of $\mathcal M$ with eigenvalue $\omega$  if and only if  $A_{\omega}  =   U_\omega  \rho_*$  for some unitary $U_\omega$ such that $M_i  U_\omega  \rho_*  =   \omega   U_{\omega} M_i  \rho_* $ for every $i  \in  \mathsf X$.
 Since $\rho_*$ is invertible, this condition is equivalent to (\ref{multiplicative}).  
 
(ii) Equation~(\ref{multiplicative}) allows us to show that if $\omega$ is an eigenvalue with eigenvector $A_\omega  =  U_\omega \rho_*$, then also its inverse $\bar \omega$  is an eigenvalue of ${\cal M}$, with eigenvector $A_{\bar \omega} :=  U_\omega^\dag \rho_*$.  
In addition, if $\omega_1$ and $\omega_2$ are two eigenvalues, with unitaries $U_{\omega_1}$ and $U_{\omega_2}$, respectively, then also $\omega_1\omega_2$  is an eigenvalue, with unitary $  U_{\omega_1}  U_{\omega_2}$ (multiplicative  rules). This proves that the eigenvalues of  $\mathcal M$ must form a group  $\mathsf F$. Clearly, the group has order $|\mathsf F|  \le d^2$, because the eigenvectors belong to the $d^2$-dimensional vector space $\mathcal B  (\mathcal H)$.    
 Moreover, since the peripheral eigenvalues lie on the unit circle, $\mathsf F$ must be a cyclic group, consisting of powers of some generator $\omega_1  =  e^{2\pi i/L}$ for some integer number $L  \le d^2$.

(iii) The multiplicative rules in the previous point imply that the operator  $A_\omega=U_\omega\rho_*$ with $U_\omega$ defined as in  (\ref{multiplicative}) is the only eigenvector of ${\cal M}$
with  eigenvalue $\omega$.  Indeed if there 
 were two of them, we would have two  unitaries $U_\omega$ and $V_\omega$ satisfying (\ref{multiplicative}). Therefore by the multiplicative properties discussed above 
 also the operator  $B =  V^\dag_\omega  U_\omega  \rho_*$  would be a fixed point for $\cal M$.  Since $\cal M$ is ergodic, we must have $  B  =  \lambda \rho_*$ for some proportionality constant $\lambda \in \mathbb C$.  Hence, $   V_\omega  = \lambda   U_\omega$. 
 
(iv) Let $A_{\omega_1} =     U_{\omega_1} \rho_* $ be the eigenvector corresponding to the eigenvalue $\omega_1  :  =e^{2\pi i/L}$.   Then, by the multiplicative rules of point (i) it follows that $A_{\omega_l} :  =U_{\omega_1}^l  \rho_* $ is an eigenvector corresponding to the eigenvalue   $\omega_l := e^{ 2\pi i l/L}$, for every $l  \in  \mathbb N$.   Since the eigenvalues are non-degenerate,  $U_{\omega_1}^l  \rho_* $ is actually  \emph{the} eigenvector   corresponding to the eigenvalue   $e^{2\pi i l/L}$ (up to a multiplicative constant).   In particular, since $\omega_L  =  1$,  we must have $U_{\omega_1}^L  =   e^{i  \alpha}     I$, for some phase $\alpha \in [0,2\pi)$. Now, without loss of generality the phase $\alpha$ can be chosen to be $0$: indeed, we can always re-defining $U_{\omega_1}$ to be  $U_{\omega_1}'  :  =  e^{i \alpha/L } U_{\omega_1}$.       
With this choice, the correspondence $\omega_l \mapsto U_{\omega_l}$ is a unitary representation of the group $\mathsf F$.   

(v) The thesis  follows by noticing that if $\bar{\omega}$ is a peripheral eigenvalue of ${\cal M}^\dag$ with eigenvector ${B}$ then Lemma~\ref{lem:strano} 
implies that ${\omega}$ must be an eigenvalue of ${\cal M}$ with eigenvector $B \rho_*$. Since ${\omega}$ is peripheral by construction and $\rho_*$ is faithful,  we can prove the thesis
by  invoking point (i) to
say that there must exist $U_{{\omega}}$ unitary such that (up to a proportionality constant) $B \rho_* = U_{{\omega}} \rho_*$. But this immediately implies that   $B$  must be proportional to $U_{{\omega}}$.

(vi) The thesis follows by noticing that 
if $\omega$ is a peripheral eigenvalue of $\cal M$ then point (i) says that its eigenvector can be written  as $A_\omega  =    U_\omega  \rho_*$ (up to a proportionality constant) with $U_\omega$ 
satisfying (\ref{multiplicative}).  From this it immediately follows that $\bar \omega$  is an eigenvalue of $ \cal M^\dag$  with eigenvector $U_\omega$:  indeed, we have ${\cal M}^\dag (U_\omega)    =  \sum_i M_i^\dag ( U_\omega   M_i)  =  \sum_i    M_i^\dag  (\bar \omega   M_i  U_\omega)   =  \bar \omega  {\cal M}^\dag  (I)       U_\omega  =  \bar \omega U_\omega$.    Note that the converse is guaranteed by Lemma~\ref{lem:strano}: if $ U_\omega$  is an eigenvector of $\cal M^\dag$ with eigenvalue $\bar \omega$, then $U_\omega  \rho_*$  is an eigenvector of $\cal M$ with eigenvalue $\omega$.
\qed

\bigskip
The fact that the peripheral eigenvalues of an ergodic channel with faithful fixed point are $L$th roots of the unit for some $L   \in  \{1, \dots,  d^2\}$ was known from \cite{WOLF} and \cite{matrixprod}.   However, the condition in terms of Kraus operators in Theorem~\ref{theo:perronfrob} allows us to prove a slightly  stronger result, namely that the peripheral eigenvalues of an ergodic random-unitary channel  are \emph{$d$-th roots} of the unit:

\begin{cor}[Peripheral eigenvalues of random-unitary channels are $d$th roots of the unit]
\label{cor:roots}
Let ${\cal M}  \in  \mathfrak C( { \cal H})$ be an ergodic channel with faithful fixed point state and let $\omega \in\mathbb C$ be a peripheral eigenvalue of $\cal M^\dag$ with $|\omega|=1$. If $\Span_{\mathbb C} \{ M_i \}_{i \in  \mathsf X}$ contains an invertible operator, then $\omega^d=1$.  
In particular, if $\cal M$ is a random-unitary channel, of the form ${\cal M} (\rho)  =  \sum_i   p_i U_i  \rho U_i^\dag$, we have $\omega^d  = 1$.     
 \end{cor}

\Proof  
Let $A_\omega  =  U_\omega \rho_*$ ($U_\omega$ unitary) be the eigenvector for the eigenvalue $\omega$.  
 Suppose that   $\Span_{\mathbb C} \{ M_i\}_{i \in  \mathsf X}$ contains an invertible operator $M$.  By linearity, the eigenvalue condition (\ref{multiplicative}) gives $ M  U  =   \omega  UM$.  Taking the determinant on both sides we obtain $ \det  (M)  \det  (U)   =   \omega^d  \det (M)  \det  (U)$.  Since $M$  and $U$ are invertible, we have $ \det  (M) \not = 0 $ and  $ \det  (U)   \not=0$. Hence, $\omega^d  = 1$. 
\qed

\bigskip
Using the fact that the peripheral eigenvalues  are roots of unit we obtain a characterization of mixing channels with a faithful fixed point. The interesting feature of this characterization is that it connects the two properties of mixing and ergodicity:

\begin{theo}[Connection between mixing and ergodicity]
Let ${\cal M}  \in  \mathfrak C( { \cal H})$ be an ergodic channel with faithful fixed point state.  The channel $\cal M$ is mixing if and only if the channel ${\cal M}^{k}$ is ergodic for all $k \le d^2$.  Moreover, if $\Span_{\mathbb C} \{ M_i\}_{ i \in  \mathsf X}$ contains an invertible operator, then $\cal M$ is mixing if and only if ${\cal M}^d$ is ergodic.      
In particular, a random-unitary channel $\cal M$ is mixing if and only if ${\cal M}^d$ is ergodic.  
\end{theo}

\Proof
If $\cal M$ is mixing, then also ${\cal M}^k$  must be mixing for every $k$, and, therefore, ergodic.   Conversely, suppose that ${\cal M}^k$ is ergodic for all $k \le d^2$ and  assume by contradiction that  ${\cal M}$ is not mixing, namely that $\cal M$ has a peripheral eigenvalue $\omega  \not  =1$ for some eigenvector $A_\omega$ which is not a multiple of $\rho_*$.   By Theorem \ref{theo:perronfrob} it follows that there exists an $L\le  d^2$ such that $\omega^L =1$. Since ${\cal M}^{L}  (A_{\omega})  =  \omega^L   A_{\omega}   =  A_\omega $, this means that ${\cal M}^L$  has two distinct fixed points $A_\omega$ and $\rho_*$, namely, it is not ergodic, in contradiction with the hypothesis.
\qed

{\blue \subsection{Ergodicity and Mixing for Unital Channels}  \label{sec:unital}}
Here we focus our attention on a particular type of channels, unital channels, which includes a large number of physically interesting examples, and allows for an even more specific characterization of ergodicity and mixing.

 We remind that a channel $\cal M \in \mathfrak C ( \cal {H})$ is called \emph{unital} if and only if it preserves the identity, that is, if and only if $\mathcal{M} (I)  =  I$.  The easiest example of unital channels is given by the class of \emph{random-unitary} channels, of the form $ {\cal M} (\rho)  =  \sum_i  p_i  U_i  \rho U_i^\dag$, with $U_i$ unitary operator and $p_i\ge 0 $ for every $i$.   Note, however, that there are many examples of unital channels that are not of the random-unitary  form \cite{UNITALQ,ALTER1,ALTER2}: as a matter of fact, as anticipated in Section~\ref{sec:subspaces} the two sets coincides only for qubit systems.

A first useful observation is that, in the case of peripheral eigenvalues, the eigenvectors of a unital channel are also eigenvectors of its adjoint:

\medskip
\begin{lemma}\label{lem:uniteigen}
Let  ${\cal M}\in  \mathfrak C ({\cal H})$  be a unital channel and $\omega \in  \mathbb C$ be a complex number on the unit circle $|\omega|=1$. 
Then $A$ is an eigenvector of ${\cal M}$ belonging to eigenvalue $\omega$ if and only if $A$ is also eigenvector  of the adjoint map ${\cal M}^\dag$ belonging to eigenvalue~$\bar{\omega}$.
\end{lemma}

\Proof
Immediate consequence of Lemma~\ref{lem:strano} and of the fact that ${\cal M}  (I)  =  {\cal M}^\dag  (I)   =  I$. 
\qed

\bigskip
\noindent
Note that unitality is an essential ingredient for the Lemma: if $\cal M$ is not unital the fixed points of $\cal M$ can easily differ from the fixed points of $\cal M^\dag$.    
For example consider the case of the erasure channel of (\ref{exera}) with $\rho_0  \not =  I/d$. While $\rho_0$ is clearly a fixed point state for  ${\cal M}$, it is not an eigenvector 
for its adjoint map ${\cal M}^\dag (A) =  I \Tr[\rho_0 A] $.

A full characterization of the peripheral eigenvalues of unital channels is the following:

\begin{theo}[Peripheral eigenvalues of unital channels]\label{theo:tuttosuunitali}  
Let ${\cal M}  \in  \mathfrak {C}(\cal{H})$ be a unital channel and $\omega \in  \mathbb C$ be a peripheral eigenvalue of $\cal M$ with $|\omega|=1$.     An operator $A  \in  {\cal B}  (\cal H)$ is an eigenvector of $\cal M$ belonging to eigenvalue $\omega$ if and only if, for every Kraus representation    ${\cal M} (\rho)  =  \sum_{i\in \mathsf X}  M_i \rho M_i^\dag$ we have 
\begin{equation}\label{eigenvector}
M_i  A  =  \omega  A  M_i, \quad \forall i \in  \mathsf X.   
\end{equation}
Moreover, the eigenspace of $\cal M$ corresponding to $\omega$ is spanned by partial isometries.
\end{theo}

\Proof
The proof is provided in the Appendix.
\qed

\subsection{Algebraic Characterization of Ergodic Unital Channels}
\begin{theo}[Characterization of ergodicity for unital channels]\label{theo:ergounit}    
For a unital channel ${\cal M}  \in  \mathfrak {C}(\cal{H})$  the followings are equivalent:
\begin{enumerate}
 \item $\cal M$ is ergodic;
\item there exists no projector $0<P< I$  such that ${\cal  M} (P)  = P$ [or equivalently, ${\cal M^\dag} (P) =  P$];
\item the matrix algebra $\cal A_{{\cal M},  {\cal M}^\dag }$ generated by the Kraus operators $\{  M_i\}_{ i \in  \mathsf X}$ and $\{  M^\dag_i\}_{i \in  \mathsf X}$ is irreducible;
\item  $\cal A_{{\cal M},  {\cal M}^\dag } =  {\cal B}  (\cal H)$;
\item if an operator $A  \in  \cal B(\cal H)$  commutes with all operators in $\cal A_{\cal M, \cal M^\dag}$ (or equivalently, with all Kraus operators and with their adjoints), then $A$ is multiple of the identity.
\end{enumerate} 
\end{theo}

\Proof  
(i) $\Rightarrow$ (ii)   If ${\cal M} (P)  = P $ for some $P< I$, then $\cal M$ cannot be ergodic because it would have two distinct fixed points  $P$ and $I$.   
(ii) $\Rightarrow$ (iii)   If $\cal A_{\cal M, \cal M^\dag}$ is reducible, then it has a proper invariant subspace $S \subset \cal H$, and denoting by $P<I$ the projector on $S$ we have $ M_i  P  =  P  M_i  P$ and $M_i^\dag P  =    P M_i^\dag P$ for every $i \in  \mathsf X$.  From these relations we get $M_i  P =  P M_i$, $\forall i$ and therefore ${\cal M}  (P)   =  \sum_i  M_i P M_i^\dag  =  \sum_i M_i M_i^\dag  P  =  {\cal M} (I)  P  = P$.  
(iii)  $\Longleftrightarrow$  (iv) $\Longleftrightarrow$ (v)  
The equivalence of (iii), (iv), and (v) is a standard fact  for operator algebras that are closed under adjoint.   
(v) $\Rightarrow$ (i)   Let $A$ be a fixed point of  $\cal M$.    By Theorem~\ref{theo:tuttosuunitali}, we have  $M_i A  =    A M_i$, $\forall i \in \mathsf X$. Since $A$ is also a fixed point of $\cal M^\dag$  we also have $M^\dag_i A  =    A M^\dag_i$, $\forall i \in \mathsf X$.      In conclusion, $A$ commutes with the Kraus operators $\{  M_i\}_{i \in  \mathsf X}$ and $\{  M^\dag_i\}_{i \in  \mathsf X}$, and hence, with all the algebra $\cal A_{M, M^\dag}$ generated by them. Hence, $A$ is a multiple of the identity.  Since every fixed point of $\cal M$ is proportional to the identity, we conclude that $\cal M$ is ergodic.    
\qed

\bigskip
Specializing to the case of random-unitary channels, Theorem~\ref{theo:ergounit} gives the interesting group-theoretic  characterization:

\begin{cor}[Group-theoretic characterization of ergodicity for random-unitary channels]
A random-unitary channel ${\cal M}   =  \sum_{i\in \sf X} p_i{\cal U}_i$ with ${\cal U}_i  (\rho)  =  U_i \rho U_i^\dag$ is ergodic if and only if the group representation generated by the unitaries $\{  U_i\}_{ i  \in  {\sf X}}$  is irreducible. 
\end{cor}

For example, the qubit channel ${\cal M}  \in  \mathfrak C (\mathbb{C}^2) $ defined by   ${\cal M} (\rho)  :=  p  X \rho X  +  (1-p)  Y  \rho Y $, where $X$ and $Y$ are Pauli matrices, is ergodic for every value $p  \in  (0,1)$.   Indeed, $X$ and $Y$ are sufficient to generate the Pauli group, which acts irreducibly on $\mathbb C^2$.  The same consideration applies in dimension $d > 2$ for the channel ${\cal M}  \in {\mathfrak C} ({\mathbb C}^d)$ defined by ${\cal  M} (\rho ) :=  p  S  \rho S^\dag +  (1-p)   M  \rho M^\dag$, where $S$ and $M$ are the shift and multiply operators, defined relative to an orthonormal basis $\{|n \rangle \}_{ n=  0}^{d-1}$ as  $  S  |n\rangle  =  |(n  + 1  )\ \mod d\rangle$  and $M  |n\rangle  =  \omega^n  |n \rangle$,  $\omega = e^{2 \pi  i/d}$, respectively.

\subsection{From Ergodicity to Mixing in the Case of Unital Channels}
We now give a sufficient condition for mixing, which has a nice algebraic form and connects mixing with ergodicity.  Unfortunately, in general this is only a sufficient condition. However, the condition is also necessary for a particular class of channels, here called \emph{diagonalizable channels}.

\begin{defi}
A channel ${\cal M} \in  \mathfrak C ({\cal H})$ is called \emph{(unitarily) diagonalizable}  if $\cal M M^\dag =  \cal M^\dag M$.
\end{defi}

The reason for the name is that a channel $\cal M$ satisfies the relation   $\cal M M^\dag =  \cal M^\dag M$ if and only if it is unitarily diagonalizable as a linear operator on ${\cal B} (\cal H)$, that is, if and only if there exists a set of complex eigenvalues $\{\mu_i\in\mathbb C\}$ an orthonormal basis for ${\cal B} {(\cal H)}$ consisting of operators $ \{  \Phi_i  \}_{i=1}^{d^2} $ such that $\langle \Phi_i, \Phi_j \rangle  =  \delta_{ij}$ and ${\cal M} (\rho)   =  \sum_{i}  \mu_i   \Phi_i \langle \Phi_i  , \rho\rangle $.  For example, all Pauli channels are diagonalizable.  
More generally, all generalized Pauli channels in dimension $d$, consisting of random mixtures of unitaries in the discrete Weyl-Heisenberg representation, are diagonalizable.

To give our condition for mixing we need to introduce the notion of \emph{square modulus} of a unital channel:

\begin{defi}
The \emph{square modulus} of  a unital channel ${\cal M} \in  \mathfrak C ({\cal H})$ is the unital channel ${\cal M}^\dag {\cal M}   \in  \mathfrak C (\cal H)$.    
\end{defi}

By definition, the square modulus is a self-adjoint non-negative operator:   $\langle A, {\cal M^\dag M}  (A)\rangle  =    \langle   {\cal M}  (A), {\cal M} (A)  \rangle  \ge 0$ for every $A  \in  {\cal B}  (\cal H)$.  This implies that $ {\cal M}^\dag {\cal M}$  can be diagonalized and has only non-negative  eigenvalues.  
  Hence, it is clear that the properties of ergodicity and mixing coincide for square moduli:  a square modulus is ergodic if and only if it is mixing.   
  Now, the main theorem of this section is the following:

\begin{theo}[Mixing from the ergodicity of the square modulus]\label{theo:mixsquare}
Let ${\cal M}  \in  \mathfrak C (\cal H)$ be a unital channel.   A sufficient condition for $\cal M$ to be mixing is that the square modulus $\cal M^\dag \cal M$ is ergodic.    If $\cal M $ is diagonalizable, then the condition is also necessary.
\end{theo}

\Proof
Suppose that the channel $\cal M^\dag M$ is ergodic.  Then, the channel $\cal M$ must be mixing.  Indeed, if an operator $A\in {\cal B (H)}$ is an eigenvector of $\cal M$ with eigenvalue $\omega$ on the unit circle, say ${\cal M} (A)  =  \omega A $,  then by  Corollary~\ref{lem:uniteigen} we have $ {\cal M}^\dag (A)   =  \bar \omega   A$ and ${\cal M^\dag M} (A)  =A$.   Since $\cal M^\dag M$ is ergodic, this implies $A  \propto I$, and therefore, $\omega =1$. In conclusion, we proved that the only peripheral eigenvalue of $\cal M$  is $\mu=1$ and is non-degenerate, i.e. $\cal M$ is mixing.  Conversely, suppose that $\cal M$ is diagonalizable and mixing.    Then, also $\cal M^\dag M$ must be mixing:  indeed, we have 
\begin{eqnarray}  
\lim_{n\to \infty}   ({\cal M^\dag \cal M} )^n  (\rho)   
 =   \lim_{n\to \infty}   {\cal M}^{\dag  n} {\cal M}^n  (\rho)
&=    \lim_{m\to \infty}      {\cal M}^{\dag m}\!\left(
 \lim_{ n\to \infty}   {\cal M}^n  (\rho)
\right)
\nonumber\\
&
=      \lim_{m\to \infty}      {\cal M}^{\dag m}(I/d)  
= I/d.      
\end{eqnarray}  
Since $\cal M^\dag M$ must be mixing, it must also be ergodic.
\qed

\bigskip
Note that the ergodicity of the square modulus is a necessary condition for mixing only in the case of diagonalizable channels. For example, consider the (non-diagonalizable) channel 
\begin{equation}
{\cal M} (\rho)  =  \sum_{n=1}^d   |e_n\rangle \langle n|  \rho |n\rangle \langle e_n| ,
\end{equation} 
where $\{|e_n \rangle\}_{n=1}^d$ is the Fourier basis ($|e_n\rangle :=  d^{-1/2}  \sum_{k=1}^de^{2\pi i  kn/d}|n\rangle$).       It is easy to see that $\cal M$ is mixing, and, in fact, that $ {\cal M}^2 (\rho)  =  I/d$ for every state $\rho$.  However, $\cal M^\dag M$ is not ergodic:  we have ${\cal M^\dag M }  (\rho)  = \sum_{n=1}^d      |n\rangle \langle n|    \rho  |n\rangle \langle n|  $ and  every projector $P_n  =  |n\rangle \langle n|$ is a fixed point of $\cal M^\dag M$.

Specializing to random-unitary channels, Theorem~\ref{theo:mixsquare} becomes:

\begin{cor}[Ergodicity of the square modulus for random-unitary channels]  
A random-unitary channel ${\cal M}  =  \sum_{i\in  \mathsf X}  p_i  {\cal U}_i$, with ${\cal  U}_i  (\rho)  =   U_i  \rho U_i^\dag$, is mixing if the group representation generated by the unitaries $\{   U_i^\dag  U_j\}_{i,j \in  \mathsf X}$ is irreducible. In particular,  a random-unitary \emph{qubit} channel $\cal M$ is mixing if the unitaries $\{   U_i^\dag  U_j \}_{ i,j \in  \mathsf X}$ do not commute.  These conditions are also necessary in the case of diagonalizable random-unitary channels.
\end{cor}

For example, while the diagonalizable qubit channel ${\cal M} (\rho)  =  p  X \rho X +  (1-p) Y\rho Y$ (denoting the Pauli matrices as $X,Y$ and $Z$) is ergodic for every $p\in (0,1)$, it is clearly not mixing because in this case the set    $\{   U_i^\dag  U_j \}_{ i,j \in  \mathsf X}$  consists of the commuting unitaries $\{  I, \pm i Z\}$.  On the other hand, the diagonalizable qubit channel $ {\cal N} (\rho)  =   p_x  X\rho X  +  p_y Y\rho Y  +  p_z Z \rho Z $ is mixing for every choice of probabilities $(p_x,p_y,p_z)  \in (0,1)^{\times 3}$.

Incidentally, we note that the ergodicity of $\cal M^\dag M$ is equivalent to a condition discussed by Streater \cite{STREATER}:

\begin{prop}\label{prop:1}
Let ${\cal M} \in \mathfrak C ({\cal H})$ be a unital channel.  The square modulus of $\cal M$  is ergodic  if and only if there is no projector $P<I$ and no unitary $U$ such that $ {\cal M} (P) =  U P  U^\dag$.
\end{prop}

\Proof
The proof can be found in the Appendix.
\qed

\bigskip
The above Proposition shows that Streater's condition is sufficient for mixing, and also necessary in the case of diagonalizable channels.

\section{Ergodic Semigroups and Ergodic Channels} 
\label{semi:SEC} 
{\red Conditions for the existence and for the uniqueness of fixed point for dynamical semigroups have been 
the subject of an intense analysis at the end of the 1970's, see e.g.\ Ref.~\cite{SPOHN} and references therein. Various characterizations of  irreducibility of the dynamical semigroup in terms of the Lindblad
decomposition were given. However, we will not review these results here. Instead, we will give an alternative characterization of ergodicity of a semigroup in terms of ergodicity of a suitable quantum channel associated to the the generator (Theorem \ref{theo:ergogroup}) 
This new characterization is useful because it allows one to translate the convexity property of ergodic channels (Theorem \ref{theo:ergoconv})  into a convexity property of  dynamical semi-groups.}

\subsection{Ergodic Channels as Generators for \emph{Mixing} CPTP Semigroups}  \label{ecgfm}
It has recently been pointed out \cite{WOLFCIR}
 that any CPTP channel ${\cal M}$ can be used to induce  a continuous CPTP semigroup evolution on $\mathfrak{S}({\cal H})$,
 described by   a  Markovian  master equation 
\begin{equation}
\label{fdds2}
\dot\rho(t) = {\cal L}^{({\cal M})} (\rho(t) )  ,  \quad \forall{t\geq 0}.
  \end{equation} 
For instance this can be done 
   by identifying 
 the generator ${\cal L}^{({\cal M})}$ of the above equation with 
the superoperator
 \begin{equation} \label{def}
{\cal L}^{({\cal M})}=\gamma({\cal M} - {\cal I}),
\end{equation} 
where $\gamma>0$ is a constant, which scales the unit of time\footnote{To see this, simply observe that ${\cal L}$  
can be cast in an explicit Lindblad form \cite{BRPE}  by introducing a set of Kraus operators $\{ M_i\}_{i\in  \mathsf X}$ for ${\cal M}$ and writing ${\cal I}(A)=
A= \frac{1}{2} \left(  \sum_{i\in\mathsf X}M_i^\dag M_i   \right) A  + \frac{1}{2} A  \left(   \sum_{i\in\mathsf X}M_i^\dag M_i  \right)$.}. It is important to stress  that the continuous  trajectories $\rho(t)$ defined by (\ref{fdds2})  in general have nothing to do with the discrete trajectories
introduced in {(\ref{CONC})}.
Indeed for the continuous case one has
$\rho(t) = {\cal T}_t^{({\cal M})} (\rho(0))$, where, for $t\geq 0$,  ${\cal T}_t^{({\cal M})}$ is an element of  the semigroup  of CPTP maps defined by 
\begin{equation} \label{ffd3}
{\cal T}_t^{({\cal M})} = e^{{\cal L}^{({\cal M})}t}  =e^{-\gamma t}  e^{\gamma t {\cal M}} ,
\end{equation} 
whose properties may be rather different from those of the  maps ${\cal M}^n$. {\blue Specific instances of dynamical semigroups of the form~(\ref{def}) have been analyzed in Ref.~\cite{KOSSAKOWSKI} (see also Ref.~\cite{SPOHN})}.
One can easily verify
that if ${\cal M}$ is ergodic, then    ${\cal L}^{({\cal M})}$ admits  a unique eigenvector associated with the null eigenvalue (up to a multiplicative factor), i.e.
\begin{equation}
 {\cal L}^{({\cal M})}(A)= 0  \quad \Longleftrightarrow \quad  A=c  \rho_*, \label{eigenvL}
\end{equation} 
where $c$ is a complex number and $\rho_*$ is the fixed point state of ${\cal M}$. Indeed from (\ref{def}) it follows that the 
 eigenvalues of ${\cal L}$ must be of the form $\mu= \gamma(\lambda-1)$ with $\lambda$ being the eigenvalues of ${\cal M}$. The condition  $\mu =0$ 
 hence implies  $\lambda=1$, which according to Lemma 1 is only possible if the eigenvector is of the form described in (\ref{eigenvL}). 
Accordingly \cite{ALICKI,SCHIRMER,BAUMGARTNER,HOLEVObook} in the limit of $t\rightarrow \infty$, the channel ${\cal T}_t^{(\mathcal{M})}$ brings all the input states toward the fixed point state of ${\cal M}$, i.e. 
 \begin{equation} \label{mixcontinuous}
\lim_{t\rightarrow \infty} \| {\cal T}_t^{({\cal M})}(\rho) - \rho_*\|_1 = 0 , \quad \forall \rho\in\mathfrak{S}({\cal H}) ,
\end{equation} 
implying that (for $t>0$) each of the maps ${\cal T}_t^{({\cal M})}$ is \emph{mixing}. 
As an example consider the case of the ergodic (but not mixing) qubit channel defined in (\ref{example}).
In this case for $n\geq 1$ integer  we have
\begin{equation}\label{example1}
{\cal M}^{2n+1}= {\cal M} ,   \qquad {\cal M}^{2n} = {\cal M}_D ,
\end{equation}  
 with ${\cal M}_D(\rho):=  \langle 0|\rho |0\rangle|0\rangle\langle 0|  +  \langle 1|\rho |1\rangle|1\rangle\langle 1| $ being the fully depolarizing channel.
Therefore,
\begin{equation}
{\cal T}_t^{({\cal M})}(\rho)= \frac{(1 - e^{-\gamma t})^2}{2} {\cal M}_D(\rho) +  \frac{1 - e^{-2\gamma t}}{2} {\cal M} (\rho)+e^{-\gamma t}\rho
\end{equation} 
which in the limit of large $t$ converges to the fixed point $(|0\rangle\langle 0| + |1\rangle\langle 1|)/2$ of ${\cal M}$, independently of the input $\rho$.

Suppose hence we have another  semigroup generator of the type defined in (\ref{def}), i.e. 
 \begin{equation} \label{def1}
{\cal L}^{({\cal E})}=\kappa({\cal E} - {\cal I}),
\end{equation} 
with ${\cal E}$ being  a (not necessarily ergodic) element of  $\mathfrak{C}({\cal H})$ and  $\kappa$ being a nonnegative constant, and consider the
evolution induced  on $S$ by the contemporary action of ${\cal L}^{({\cal M})}$ and ${\cal L}^{({\cal E})}$, i.e. 
\begin{equation}
\label{fdds22}
\dot\rho(t) = ({\cal L}^{({\cal M})}+ {\cal L}^{({\cal E})})(\rho(t)),  \quad \forall{t\geq 0}.
  \end{equation} 
 Introducing the parameters  $\tilde{\gamma} = \gamma+\kappa$ and $\lambda = \gamma/\tilde{\gamma}\in {(0,1]}$ we  observe that the resulting
 Lindblad superoperator can be cast again in the form (\ref{def}), for the CPTP map $\tilde{\cal M} = \lambda {\cal M} + (1-\lambda){\cal E}$, i.e.  
 \begin{equation} 
   {\cal L}^{({\cal M})}+ {\cal L}^{({\cal E})} =   {\cal L}^{(\tilde{\cal M})}  = \tilde{\gamma}  
   (\tilde{\cal M} - {\cal I}).
   \end{equation} 
According to the  results of the previous section we can then conclude that the continuous trajectory 
${\cal T}_t^{(\tilde{\cal M})}(\rho)$ associated with (\ref{fdds22}) 
is again mixing, independently of the ratio $\lambda$ and of the properties of ${\cal L}^{({\cal E})}$ (the asymptotic 
convergence point being the fixed point of the ergodic channel $\tilde{\cal M}$).

\subsection{Necessary and Sufficient Condition for the Ergodicity of a Lindblad Generator} 
A generalization of the result discussed in the previous section to arbitrary Lindblad generators 
can be obtained by reversing the connection ${\cal M}\rightarrow {\cal L}^{({\cal M})}$ of (\ref{def}). Specifically we will show  
that the ergodicity of a  \emph{generic}  Lindblad generator ${\cal L}$ (and hence the asymptotic mixing property of its integrated trajectory $e^{{\cal L}t}$) is   
 equivalent to the ergodicity of a suitable quantum channel $\cal M_L$ which can be  associated to~${\cal L}$.  
 To see this we recall that any ${\cal L}$ can always be written~as 
\begin{eqnarray}
 {\cal L}  (\rho)  
&=   i [H, \rho]  +   2 {\cal A}  (\rho)  -      {\cal A}^\dag  (I)  \rho  -  \rho  {\cal A}^\dag  (I) \nonumber\\ 
&=    2 {\cal A}   (\rho) - G \rho  -  \rho G^\dag,   \quad
G :  =   {\cal A}^\dag (I)    -  i H,      
\label{completely dissipative}
\end{eqnarray}
where  $H = H^\dag$ is an Hamiltonian operator and  
$\cal A$  is a completely positive (not necessarily trace preserving) map.     
Equation~(\ref{completely dissipative})  can be conveniently rewritten as a difference of two completely positive maps:     
\begin{eqnarray}
{\cal L}  =   {\cal   S_L}  -  {\cal T_L},  \label{deflll}\\
     {\cal S_L}    (\rho)  : =  2  {\cal A}  (\rho)  +   \frac 1 2    (I  - G)        \rho      ( I  - G^\dag  ),  \\
 {\cal T_L}  (\rho) :  =   \frac 1 2    ( I  +   G   )       \rho    (  I +  G^\dag )  .
\end{eqnarray} 
where the second term is invertible with completely positive inverse 
\begin{eqnarray}
{\cal T}_{\cal L}^{-1}(\rho)  :=  \frac 1 2    ( I  +   G   )^{-1}       \rho    (  I +  G^\dag )^{-1} ,
\end{eqnarray}
[the invertibility of ${\cal T_L}$ being a direct consequence of the invertibility of the operator $I+G$, the latter following from the fact that 
 ${\cal A}^{\dag}(I)$ is non-negative].
 The definition of the quantum channel ${\cal M_L}$  is now obtained by observing that

\begin{lemma}  
The map ${ \cal M}_{\cal L}  :  =     { \cal S_L} \circ {\cal T_L}^{-1}  $ is completely positive and trace-preserving.  
   \end{lemma}

\Proof
Complete positivity is clear.  To prove that $\cal M_L$  is trace-preserving   we show that its dual ${\cal M_L}^\dag$ is unit-preserving.  Indeed, we have  
 \begin{eqnarray}
{\cal M}_{\cal L}^\dag  (I )  
   &=    2  ( I + G^\dag )^{-1}     \left[  
   2   {\cal A}^\dag  (I)    +    \frac 1 2    (I  - G^\dag) (I-G)         
   \right ]        (I-G)^{-1}\nonumber  \\  
   &  =     2(I + G^\dag)^{-1}     \left[   
   G +  G^\dag  +  \frac 1 2    (I -  G^\dag) (I- G)         
   \right ]         (I+G)^{-1} \nonumber\\
   &  =     2   ( I + G^\dag )^{-1}     \left[   
   \frac 1 2    (I +  G^\dag) (I + G)         \right ]         (I+G)^{-1} 
   =  I.
 \end{eqnarray}
\qed

\begin{theo}[Ergodic semigroups and ergodic channels]\label{theo:ergogroup}  
 The semigroup generated by $\cal L$ in (\ref{completely dissipative}) is ergodic if and only if the channel $\cal M_L$  is ergodic.  
 \end{theo}

\Proof  
Let $A$ be a fixed point of the semigroup generated by $\cal L$, namely $  {\cal L}  (A)  =  0$.   
 Equivalently, we have   $  {\cal S_L}  (A)  =   {\cal T_L}  (A)$ which implies  that ${\cal T_L}  (A)$  is a fixed point of the channel ${\cal  M_L}  = {\cal S_L}  \circ {\cal  T}^{-1}_{\cal L}  $.   Hence, the semigroup generated by $\cal L$  has a unique fixed point if and only if $\cal M_L$ has a unique fixed point. 
\qed

\bigskip
As a special instance of the Theorem, we can re-obtain the results of Section~\ref{ecgfm}. Indeed 
 assuming ${\cal L}$ as in (\ref{def}) we have  $H  =   0$,  ${\cal A}  = (\gamma/2) {\cal M}$ with ${\cal M}$ being CPTP so that $  G  =  {\cal A}^\dag (I)  =  \gamma I/2$.    Hence, the maps $\cal S_L$, $\cal T_L$, and $\cal M_L$ are given by ${\cal S_L } = \gamma {\cal  M}  +  \frac{(2-\gamma)^2}{8}{\cal I}$,  ${\cal T_L }=    \frac{(2-\gamma)^2}{8}  \cal I $, and  ${\cal M_L}  =   
   \frac{8 \gamma }{(2+\gamma)^2}   {\cal M}     +  \frac{(2-\gamma)^2 }{(2+\gamma)^2}   \cal I  $.   Note that $\cal M_L$ (and hence ${\cal L}$) is ergodic if and only if $\cal M$  is.

\subsection{Convexity of Ergodic Semigroups} \label{sec:semi2}
Consider now two dynamical semigroups, with Lindblad generators $\cal L$ and $\cal L'$ 
and define the Lindblad generator  
\begin{equation}
{\cal L}_p :  =  p  {\cal L}  +  (1-p)  {\cal L'}, \quad p  \in   (0,1].
\end{equation}   
Assuming that $\cal L$ generates an ergodic semigroup, we may ask whether ${\cal L}_p$ also generates an ergodic semigroup.  At the end of  Section~\ref{ecgfm} we have
already seen that this is indeed the case when~(\ref{def}) hold for both generators. 
Answering to the  question for the general case is difficult. Still it is possible to provide a relatively simple answer 
 when, cast in the form~(\ref{completely dissipative}),  the two semigroups 
have the same Hamiltonian and the same positive operators ${\cal A}^\dag  (I)$  and ${\cal A}^{\prime\dag} (I)$, that is,
\begin{equation}
H'    =      H,\qquad
{\cal A}^{\prime\dag}  (I)   =      {\cal A}^{\dag}  (I)  \label{prop}
\end{equation}  
(incidentally this case covers also the scenario addressed in Section~\ref{ecgfm}).

\begin{theo}[Convexity of ergodic semigroups]\label{convexitysemigroup}  
Suppose that the condition of (\ref{prop})  is satisfied and that $\cal L$  generates an ergodic semigroup. Then,  ${\cal L}_p$ generates an ergodic semigroup for every $p\in  (0,1]$. 
\end{theo}

\Proof  
Under the condition of (\ref{prop}), we have 
\begin{eqnarray}
 {\cal S}_{{\cal L}_p}(\rho) 
 &=p   {\cal A} (\rho)  +   (1-p){\cal A}'  (\rho)  +  \frac 12  (I-G)  \rho    (I-G^\dag)     \nonumber\\
 &=p  {\cal S}_{{\cal L}}  (\rho)+  (1-p)   {\cal S}_{{\cal L}'} (\rho), \\
{\cal T}_{{\cal L}_p}(\rho) &=\frac 12  (I+ G)  \rho    (I+G^\dag)   =  {\cal T}_{{\cal L}} (\rho) = {\cal T}_{{\cal L}'}  (\rho),  
\end{eqnarray}
and therefore,   ${\cal M}_{{\cal L}_p}  =  p   {\cal M}_{{\cal L}}  +(1-p)  {\cal S}_{{\cal L}'}$.   Using Theorems~\ref{theo:ergoconv} and~\ref{theo:ergogroup} we then obtain that ${\cal L}_p$ is ergodic.
\qed

\bigskip
Because of condition (\ref{prop}), this is a weaker Theorem than the corresponding Theorem~\ref{theo:ergoconv} for discrete channels. It is easy to see that generally, ergodicity is not stable under convex combination. As a simple example, consider the qubit Lindbladians ${\cal L}_\pm(\rho) = \pm i[X,\rho] + (Z \rho Z - \rho)$. As they contain dephasing to the Z-axis combined with a rotation around the X axis, they are easily seen to be ergodic, with the centre of the Bloch sphere as fixed point. However, their midpoint convex combination ${\cal L}=({\cal L_+}+ {\cal L_-})/2$ is simply a dephasing map, which leaves the whole Z-axis invariant.

Despite being weaker than Theorem~\ref{theo:ergoconv}, we can still conclude that the set of non-ergodic dynamical semigroups have measure zero. Note that the condition (\ref{prop}) is equivalent to $G=G'$. We decompose the set of all Lindblad superoperators into convex sets with $G=G'$. Now, each of this set contains at least one ergodic Lindblad superoperator: simply choose $ {\cal A} (\rho)  =     \rho_*    \Tr[  \rho ( G + G^\dag)/2 ]$, where $\rho_*$ is faithful. The channel  $\mathcal{M}$ associated to this Lindbladian is of the form    $\mathcal{M} (\rho)  =    \rho_*       \Tr[  P \rho  ]    +  \mathcal{Q}  (\rho)$,
where $P$ is a positive operator and $\mathcal{Q}$ is a quantum operation. A channel of this form is necessarily ergodic, due to our characterization Theorem~\ref{theo:ergogen}, because $\mathcal{M}$ cannot have two  distinct invariant subspaces. Therefore, the only non-ergodic maps can be at the boundary of the sets with $G=G'$.

\section{Conclusions} \label{sec:CON}
We have discussed structural properties of quantum channels in finite dimensions, focusing on criteria for ergodicity and mixing. Because these notions are relevant to many protocols in quantum information processing, our characterization paves the way to simpler proofs of quantum convergence in those applications. One of our main results, i.e.\ the convexity of ergodicity, has potential applications for toy models \cite{ZIMAN, MEDIATED} in quantum statistical dynamics, where the ergodicity of a given model is usually hard to establish. Since thermal states provide a natural convex decomposition, implying a convex decomposition of the corresponding map, our result implies that it suffices to establish ergodicity at zero temperature only. 

\appendix
\section{}
This section is dedicated to the detailed discussions of  some of the technical aspects we presented in the main text. 
We start by presenting a useful Lemma which is often invoked in the text.

\begin{lemma} \label{lemma:sup} 
Given  $\rho,\sigma \in  \mathfrak{S}({\cal H})$ density matrices, they can be related as
\begin{equation}
\rho= q \sigma + (1-q) \tau, \label{convconv}
\end{equation} 
  with $q\in(0,1]$ and  $\tau \in  \mathfrak{S}({\cal H})$, if and only if  $\Supp(\sigma) \subseteq \Supp(\rho)$.
  \end{lemma}

\Proof
Consider first the case in which  (\ref{convconv}) holds for some $q$ and $\tau$. By contradiction assume then that 
$\Supp(\rho) \subset \Supp(\sigma)$, i.e.\ $\Ker(\sigma) \subset  \Ker(\rho)$, where
$\Ker(A)$ represents  the kernel of a self-adjoint operator $A$. 
This implies that one  can identify a vector $|\psi\rangle$ such that  $|\psi\rangle \in \Ker(\rho)$ and  $|\psi \rangle \not\in  \Ker(\sigma)$. For it we must
have 
\begin{equation} 
0 =  \langle \psi|\rho|\psi\rangle  =  q \langle \psi |\sigma |\psi\rangle  + (1 - q)\langle \psi|\tau  |\psi\rangle  > 0,
\end{equation}  
which  is clearly absurd, and the necessity is proved.  On the contrary, suppose   $\Supp(\sigma) \subseteq \Supp(\rho)$. Let us take the maximum eigenvalue $\lambda$  of $\sigma$ 
and the minimum non-null eigenvalue $\mu$ of $\rho$. Clearly $\rho \geq \mu P_\rho$ and $\sigma \leq \lambda P_\rho$, where 
$P_\rho$ is the   projection on the
 support of $\rho$. It is possible to prove that 
$\mu\leq \lambda$ since $\mu \leq 1/ r_{\rho} \leq 1 /r_{\sigma} \leq \lambda$,
with $r_A$ denoting the rank of a generic operator $A$.  
Now if  in the previous relation we have an equality then $\rho = \sigma = P_\rho/r_\rho$ and the thesis derives immediately. 
 On the other hand, if $\mu < \lambda$ then  we have that $(\mu/\lambda) \sigma \leq \mu P_\rho \leq \rho$, which implies   $\rho=(\mu/\lambda) \sigma + (1 -\mu/\lambda ) \tau$, with  $\tau\geq 0$. But this is exactly of the form (\ref{convconv}):  indeed  since $\mu < \lambda$  and  $\Tr[\rho]=1$, we have 
 $(1- \mu/\lambda) \Tr [\tau]= 1 -\mu/\lambda$, which implies that $\Tr[\tau]=1$ and hence $\tau  \in  \mathfrak{S}({\cal H})$. 
\qed

\subsection{Other Proofs} 
Here we conclude the Appendix by providing the details of the proofs that had been skipped in the main text.

\medskip
\noindent
\Proofa{Proof of Lemma \ref{lem:positivefix}}
Since  $A$  is a fixed point, also $A^\dag$ is a fixed point [see (\ref{FF11})] and so are the linear combinations  $X  :=  (A  + A^\dag)/2$  and $Y  :  =  (A-A^\dag)/2i$.   Let us denote by $P_+$ ($P_-$) the projectors on the eigenspaces of $X$ with non-negative (negative) eigenvalues and write $X$ as $X  =  X_+  - X_-$ where $X_+ :=  P_+  X $ and $X_-:  =  - P_- X $.   Now, we have $X_+  = P_+  X  =   P_+  {\cal M}  (X)   =  P_+  {\cal M}  (X_+)  -  P_+  {\cal M}  (X_-)    $, which implies 
\begin{eqnarray}
\Tr[X_+]  &  =  \Tr [P_+  {\cal M} (X_+)]  -   \Tr [P_+  {\cal M} (X_-)]\nonumber\\
&  \le    \Tr [P_+  {\cal M} (X_+)]
  \le    \Tr [ {\cal M} (X_+)]
=    \Tr[X_+].
\end{eqnarray}  
In order for the equality to hold, it is necessary to have $P_+  {\cal M}  (X_-)  = P_-  {\cal M} (X_+)=0$. Hence, we have $X_+  =  P_+  {\cal M} (X_+)  =  {\cal M} (X_+)$, which also implies $X_-  =  {\cal M} (X_-)$.   Repeating the same reasoning for $Y  = Y_+  - Y_-$ we obtain that also $Y_+$ and $Y_-$ are fixed points of $\cal M$. 
\qed

\bigskip
\noindent
\Proofa{Alternative Proof of Corollary~\ref{lem:unique}}
  From Lemma 6 of \cite{NJP},
  we know that $|A|:=\sqrt{A^\dag A}$  must be a fixed point state of the map. 
 If ${\cal M}$ is ergodic with fixed point state $\rho_*$ then we must have $|A|=\|A\|_1\rho_*$.
Consider first the case in which $A$ is Hermitian, i.e.\ $A =A^\dag$ (the non-Hermitian case will be considered below). 
In this case we can then write 
$\rho_*= \sum_n|c_n|\ket{\phi_n}\bra{\phi_n}/\sum_m |c_m|$,
with $\{ |\phi_n\rangle\}$ being the orthonormal eigenvectors of $A$ and $c_n$ the corresponding eigenvalues.
Furthermore for each $\alpha$ real  and satisfying the inequality
$0<|\alpha|\le1/\sum_m|c_m|$,
we can also conclude that  the operator 
\begin{equation}
\tilde{\rho}_*
=
 \frac{\alpha A + \rho_*}{\alpha \Tr[A] + 1} 
 =
 \frac{1}{\alpha\sum_m c_m+1}
  \sum_n\left(
\alpha c_n+\frac{|c_n|}{\sum_m|c_m|}
\right)
\ket{\phi_n}\bra{\phi_n}
\end{equation}
is  a density matrix of the system (indeed it has trace $1$ and is positive semidefinite) which 
by construction  is also a fixed point of $\mathcal{M}$.
Therefore, since $\mathcal{M}$ is ergodic, we must have
 $\tilde{\rho}_* =\rho_*$ or  $A = \rho_*\Tr[A]$,
as required by the Corollary~\ref{lem:unique}. 
Suppose now that $A$  is not Hermitian. Since $A^\dag$ is also an eigenoperator
with eigenvalue $1$, the Hermitian operators $(A+A^\dag)/2$ and 
$(A - A^\dag)/2i$ are also solutions of (\ref{eq2}).
Hence, from the previous derivation, we must have 
$(A+A^\dag)/2 = \rho_*\Tr[A+A^\dag]/2$ and 
$(A-A^\dag)/2i = \rho_*\Tr[A-A^\dag]/2i$,
which yields (\ref{eq3}). 
\qed

\bigskip
\noindent
\Proofa{Proof of Theorem~\ref{theo:tuttosuunitali}}
If (\ref{eigenvector}) holds, then we have ${\cal  M} (A)  =  \sum_i  M_i A M_i^\dag =  \omega A   \sum_i  M_i  M_i^\dag   = \omega A  {\cal M} (I)  =  \omega A$.    Hence,  $A$ is an eigenvector of $\cal M$ with eigenvalue $\omega$.    
Conversely, suppose that $A$ is eigenvector belonging to eigenvalue $\omega$ with $|\omega| =1$.  Equivalently, we have $  {\cal M}^\dag (A)  = \bar \omega A$ (cf.\ Corollary~\ref{lem:uniteigen}).    Writing $A$ in the polar decomposition $A  =     U  |A|$ and choosing a Kraus form for $\cal M$  we have 
 \begin{equation}
 |A|  =    \omega \sum_i     (U^\dag M^\dag_i  U)  |A|   M_i.           
 \end{equation}  
 Let us diagonalize $|A| $ as $|A|  =  \sum_{k=1}^s   \alpha_k    P_k$, where $\alpha_1 > \alpha_2  > \dots > \alpha_s\ge 0$ are the eigenvalues, $S_k$ is the  eigenspace corresponding to $\alpha_k$, and $P_k$ is the projector on $S_k$.
Then, for every unit vector $\varphi  \in  S_1$ we have  
\begin{eqnarray}
\alpha_1 
&= \langle \varphi | \, |A|\,  |\varphi \rangle
=  \sum_i    \omega \langle \varphi |        U^\dag M_i^\dag  U |A|   M_i    |\varphi \rangle \nonumber\\  
&\le  \sqrt{   \sum_i       \langle \varphi |U^\dag M^\dag_i  U  |A|U^\dag M_i  U    |\varphi \rangle\cdot\sum_j       \langle \varphi |       M_j^\dag   |A|    M_j     |\varphi \rangle}, \label{schwarz}
\end{eqnarray}  
where the inequality comes from the Cauchy-Schwarz inequality.  
To saturate the inequality (\ref{schwarz}) we need to have
\begin{equation}\label{needed}
 \exists  \lambda \ge 0
 \ \ \mathrm{s.t.}\ \ %
 |A|^{1/2} U^\dag M_i  U  |\varphi \rangle   =  \lambda   \omega  |A|^{1/2}  M_i     |\varphi \rangle,\ \ \forall i  .  
\end{equation}    
Continuing the inequality (\ref{schwarz}) we then get
\begin{equation}
\alpha_{1}
\le  \alpha_{1}  \sqrt{  \sum_i       \langle \varphi |      (U^\dag M_i^\dag  U)    (U^\dag M_i  U)    |\varphi \rangle \cdot\sum_j       \langle \varphi |       M^\dag_j       M_j     |\varphi \rangle} 
=   \alpha_{1}  .
 \label{invsub}
\end{equation}
Therefore, both  inequalities (\ref{schwarz}) and (\ref{invsub}) must be saturated.  To saturate the  inequality (\ref{invsub}) it is necessary to have $ M_i  |\varphi\rangle  \in S_1$ and $ U^\dag  M_i U |\varphi\rangle  \in S_1$ for every $i$, so that (\ref{needed}) becomes 
\begin{equation}\label{almost}
\exists \lambda \ge 0
\quad\mathrm{s.t.}\quad
U^\dag M_i  U  |\varphi \rangle   =  \lambda    \omega      M_i     |\varphi \rangle, \quad \forall i .   
\end{equation}  
Clearly, in order for $\cal M$ to be trace-preserving we must have $\lambda=1$, as $\lambda^2  =      \sum_i  \langle \varphi|    (\lambda \bar \omega M^\dag_i )(\lambda  \omega  M_i)  |\varphi \rangle  = \sum_i  \langle \varphi| (U^\dag M_i^\dag  U)  (U^\dag M_i  U) |\varphi\rangle=1$.  Moreover, since   $\ket{\varphi}$ is a generic element of $S_1$,  (\ref{almost}) with $\lambda = 1$ is equivalent to
\begin{equation}\label{quasi}
  (U^\dag M_i  U)  P_1      =    \omega       M_i   P_1,  \quad \forall i .   
\end{equation}

Similarly, the relation $M_i  |\varphi\rangle  \in  S_{1}$, $\forall\ket{\varphi }\in  S_{1}$, which was needed to saturate the inequality (\ref{invsub}),  is equivalent to 
\begin{equation}\label{uno}
M_i P_1  =  P_1  M_i   P_1,  \quad \forall i.
\end{equation}  
Recalling that the eigenvalue equation ${\cal M}^\dag(A)  = \bar  \omega  A$  is equivalent to ${\cal M} ( A) =  \omega A$ (Corollary~\ref{lem:uniteigen}), we can use the same reasoning as above to prove also 
\begin{equation}\label{due}
M_i^\dag P_1  =  P_1  M_i^\dag   P_1,  \quad \forall i.
\end{equation}  
Putting together the two relations (\ref{uno})  and (\ref{due}) we then obtain $M_i P_1  =  P_1  M_i$, $\forall i$.
Finally, defining the partial isometry $T_1 :  =    U  P_{1}$ we obtain
\begin{eqnarray}
 M_i   T_1 &=     M_i  U  P_1  =   U (U^\dag M_i U)  P_1  \nonumber\\
  &= U  (\omega  M_i P_1) =  \omega  (UP_1)  M_i   =  \omega T_1  M_i , \quad \forall i.          
\end{eqnarray}
Hence, we proved that the partial isometry $T_1$ must satisfy (\ref{eigenvector}) in the statement of the theorem.  In particular, we then have ${\cal M} (T_1)  =  \omega T_1$.    This means that for every peripheral eigenvalue $\omega$, the channel $\cal M$ must have at least one eigenvector that is a partial isometry and satisfies (\ref{eigenvector}).   
Moreover, defining the operator $A' :=A  -  \alpha_{1}   T_1$  we have ${\cal M} (A')  =  \omega A' $.  The polar decomposition of $A'$ is $A' =  U (  |A|  -  \alpha_{1}   P_{1})   =  U  \sum_{k=2}^s    \alpha_k   P_k$, so that the eigenspace of $|A'|$ with maximum eigenvalue is  $S_2$.   Iterating the above proof we obtain that the partial isometry $T_2: = U P_2$  is an eigenvector of $\cal M$ satisfying (\ref{eigenvector}), and by further iteration we obtain that every partial isometry $T_k :=  U P_k$  is an eigenvector of $\cal M$ satisfying (\ref{eigenvector}).   In conclusion, the operator $A$ is a linear combination of partial isometries satisfying    (\ref{eigenvector}) and, by linearity, it satisfies  (\ref{eigenvector}).  
\qed

\bigskip
\Proofa{Proof of Proposition~\ref{prop:1}}
Suppose that $\cal M^\dag M$ is not ergodic.  Hence, by {Theorem~\ref{theo:ergounit}} there exists a projector $P  < I$  such that $({\cal M^\dag M})  (P)  =  P$.   Using this fact we obtain  
\begin{equation}
 \langle  {\cal M}  (I-P) ,   {\cal M}  (P)  \rangle  =  \langle     I-P  ,  {\cal M^\dag \cal M}   (P)  \rangle
 =    \langle  I-P ,   P  \rangle  =  0.  
\end{equation}  
Since $\Tr [  {\cal M}  (I-P)    {\cal M}  (P) ]=0$, we necessarily have $ {\cal M}  (I-P)    {\cal M}  (P) =0$.    In addition we have $   {\cal M}  (I-P) +   {\cal M}  (P)   =  {\cal M}  (I)  =  I$.   It is then easy to see that  $ {\cal M}  (P) $ is an orthogonal projector:  we have 
\begin{equation}
[{\cal M}  (P)]^2 
 =    {\cal M}   (P)   [ { \cal M}(P)  +  {\cal M} (I-P) ]   
=  {\cal M}  (P)   {\cal M}  (I)  =  {\cal M} (P).       
 \end{equation}  
Since we have $\Tr  [{\cal M}  (P) ] = \Tr  [  P ]$, the dimensions of the support of ${\cal M} (P)$ and  of $P$  coincide. Hence, there exists a unitary $U$ such that $  {\cal  M} (P)  =  U  P  U^\dag$.  
 Conversely, suppose that the relation  ${\cal M} (P)  =  U  P U^\dag$ is satisfied for some projector $P<I$ and for some unitary $U$.   Then, we have 
\begin{eqnarray}
\Tr  [P]& =    \langle  U P  U^\dag,   {\cal M} (P)  \rangle 
    =   \langle  {\cal M}^\dag (U P  U^\dag),    P  \rangle \nonumber\\
    & \le \sqrt{   \langle  {\cal M}^\dag (U P  U^\dag),     {\cal M}^\dag (U P  U^\dag) \rangle     \langle  P,     P  \rangle }\nonumber\\   
     & =  \sqrt{  \Tr [U P  U^\dag        {\cal M  M}^\dag (U P  U^\dag) ] \Tr [ P] }\nonumber\\     
     & \le \sqrt{ \Tr [   {\cal M  M}^\dag (U P  U^\dag) ]     \Tr [ P] }
=   \Tr [P],            
\end{eqnarray} 
which requires $  {\cal M}^\dag (U P  U^\dag) = P  $.    Hence, we have $  {\cal M^\dag  \cal M} (P)  =  {\cal M^\dag}   (UPU^\dag)  =  P $, namely $P$ is a fixed point of $\cal M^\dag M$. Hence,   $\cal M^\dag M$ is not ergodic.
\qed

\ack
We acknowledge discussions with David Reeb, Kai Schwieger, and Alexander S Holevo.
This work is partially supported by the Joint Italian-Japanese
Laboratory on “Quantum Technologies: Information, Communication and
Computation” of the Italian Ministry of Foreign Affairs.
GC acknowledges support by the National Basic Research Program of
China (973) 2011CBA00300 (2011CBA00301), the National Natural Science
Foundation of China through Grants 11350110207, 61033001, and
61061130540, the 1000 Youth Fellowship Program of China, and the kind
hospitality of the Institute of Theoretical Computer Science and
Communications, Chinese University of Hong Kong.
VG acknowledges support from   MIUR through FIRB-IDEAS Project  No.\ RBID08B3FM\@.
KY is supported by the Grant-in-Aid for Young Scientists (B) (No.\ 21740294) and the Grant for Excellent Graduate Schools both from the Ministry of Education, Culture, Sports, Science and Technology (MEXT), Japan, and by a Waseda University Grant for Special Research Projects (2013B-147).

\section*{References}

\end{document}